\documentclass[twocolumn,twocolappendix]{aastex631}
\usepackage{amsmath,amssymb}
\usepackage{bm}

\shorttitle{SRF Catalysis of Extreme Orbital Evolution}
\shortauthors{Klein \& Hamilton}

\newcommand{\p}{\partial}

\newcommand{\ba}{\begin{eqnarray}}
\newcommand{\ea}{\end{eqnarray}}

\newcommand{\emax}{e_{\rm max}}
\newcommand{\emin}{e_{\rm min}}

\newcommand{\Psec}{P_{\rm sec}}
\newcommand{\Pout}{P_{\rm out}}

\newcommand{\epsGR}{\varepsilon_{\rm GR}}
\newcommand{\epsTide}{\varepsilon_{\rm Tide}}
\newcommand{\epsRot}{\varepsilon_{\rm Rot}}

\newcommand{\CK}{C_\mathrm{K}}

\defcitealias{hamilton2024}{HR24}

\providecommand{\rev}[1]{#1}
\providecommand{\revcite}[1]{\citep{#1}}

\makeatletter
\AtBeginDocument{}
\makeatother

\begin{document}


\title{Short-Range Forces Can Catalyze Extreme Orbital Evolution\\
in Hierarchical Triples}

\correspondingauthor{Ygal Y. Klein}
\email{ygalklein@gmail.com}

\author[0009-0004-1914-5821]{Ygal Y. Klein}
\affiliation{Institute for Advanced Study,
Einstein Drive, Princeton, NJ 08540, USA}

\author[0000-0002-5861-5687]{Chris Hamilton}
\affiliation{Department of Astrophysical Sciences, Princeton University,
4 Ivy Lane, Princeton, NJ 08544, USA}
\affiliation{Institute for Advanced Study,
Einstein Drive, Princeton, NJ 08540, USA}
\email{chamilton@princeton.edu}


\begin{abstract}
Hierarchical triples are promising environments for producing exotica such as black hole mergers and hot Jupiters, because of the
von Zeipel-Lidov-Kozai (ZLK) effect, whereby a distant tertiary can torque an inner binary to high eccentricity over secular timescales. In the double-averaged (DA) approximation to ZLK, this eccentricity excitation is
suppressed by apsidal precession due to `short-range forces' (SRFs) like relativity and tidal/rotational bulges.  
Here we show that, \rev{when the DA approximation breaks down,} SRFs often catalyze, rather than suppress, extreme
eccentricity behavior. This occurs because SRFs can drive large, discrete jumps in the binary's effective `adiabatic invariants' during
high-eccentricity episodes. 
These nonadiabatic jumps can dramatically alter the maximum/minimum eccentricity and secular period of astrophysically relevant triples, including some for which SRFs were previously thought irrelevant.
Even the angular momentum component $j_z$ evolves secularly --- to our knowledge, this is the first time such evolution has been demonstrated from a \rev{test-particle} quadrupole-order, three-body mechanism. In short, binaries may explore much more of phase space than is implied by any (semi-)analytic ZLK theory of which we are aware.
We demonstrate this at the test-particle quadrupole level; in a companion \rev{work} we show how even more-extreme behavior occurs when the jumps are combined with octupolar ZLK evolution.
\end{abstract}


\section{Introduction}
\label{sec:intro}

Hierarchical triple systems are widely studied as potential origin sites of exotic astrophysical objects because of the famous von Zeipel-Lidov-Kozai (ZLK) oscillations \citep{vonzeipel1910,lidov1962,kozai1962}\footnote{For a historical overview, see
\citet{ito2019}.}. Precisely, an inclined tertiary companion can periodically torque an inner binary to very high eccentricity (i.e., to very small pericenter distance) on secular timescales. At small pericenter distances, interactions between the binary components can occur that would not have been possible in the absence of the ZLK effect, such as dissipation through gravitational wave emission or tides, or even head-on collisions. This has led many authors to identify ZLK oscillations as the culprit behind the formation of (at least some) hot Jupiters and the eccentricity distribution of cold Jupiters \citep{naoz2011,naoz2012,petrovich2015,anderson2016,munoz2016,owen2018,vick2019,teyssandier2019,oconnor2021,stephan2021,angelo2022,stephan2024wdkick,weldon2025,grishin2026}, binary black hole (BBH) mergers \citep{miller2002,blaes2002,wen2003,antonini2012,naoz2013pn,antonini2016,stephan2016,liu2017,hoang2018,grishin2018,liu2018,hamilton2019a,liu2019quad,martinez2020,fragione2019,vignagomez2025}, neutron-star gravitational-wave sources \citep{feng2026}, blue stragglers \citep{perets2009,naoz2014,antonini2016}, tidal disruption events \citep{ivanov2005,melchor2024,melchor2025}, Type Ia supernovae \citep{thompson2011,katz2012}, white-dwarf binaries, cataclysmic variables (CVs), and low-mass X-ray binaries (LMXBs) \citep{shariat2023wd,shariat2025cv,shariat2025lmxb}, and the formation of Kuiper Belt contact binaries such as Arrokoth \citep{grishin2020}.

Whatever the astrophysical context, a theory of ZLK dynamics needs to answer two basic questions:
\begin{itemize}
    \item[] (i) what is the maximum possible eccentricity $e_\mathrm{max}$ that the inner binary can reach? and 
    \item[] 
    (ii) how long does it take to get there, i.e., what is the secular period (or `Kozai period') $P_\mathrm{K}$? 
    \end{itemize}
The simplest theory that answers these questions rests on certain physical and mathematical approximations.  It assumes pure Newtonian gravity, with no short-range forces (SRFs) from general relativity, tides, or rotational bulges, and no dissipation; it treats the inner binary in the `test-particle' approximation so the outer orbit feels no back-reaction from it. Then the tertiary's tidal potential is expanded in powers of the inner-to-outer semimajor axes ratio $a/a_2 \ll 1$, truncated at the leading non-trivial (quadrupole) order, and the resulting equations are double-averaged (DA) over both the inner and the outer orbit. In this limit the dynamics preserves two integrals of motion which, in dimensionless form, can be taken as the angular momentum component perpendicular to the outer orbital plane, $j_z \in [-1, 1]$, and the averaged Hamiltonian, or `Kozai constant', $C_\mathrm{K} \in [-3/2, 1]$. This setting is completely integrable \citep{lidov1962,kozai1962} and is in fact equivalent to that of a simple pendulum \citep{basha2025}. Both questions (i) and (ii) can be answered precisely from just the two numbers $j_z$ and $C_\mathrm{K}$; the maximum eccentricity $e_\mathrm{max}$ is mostly set by $j_z$ (with smaller $\vert j_z \vert$ giving higher $e_\mathrm{max}$), and the secular period $P_\mathrm{K}$ is mostly set by $C_\mathrm{K}$ (with smaller $\vert C_\mathrm{K} \vert$ giving longer $P_\mathrm{K}$). This \textit{Newtonian, test-particle, quadrupole, DA} setting is the simple, integrable baseline against which more realistic versions of the ZLK problem are usually analyzed.

It has long been known that the answer to question (i), the maximum eccentricity reached by the inner binary, changes somewhat when one modifies the assumptions of this simple baseline case. Meanwhile, the answer to question (ii), the secular period $P_\mathrm{K}$, was thought to be essentially robust to these changes.   The purpose of this paper is to demonstrate that, in various realistic scenarios, the (semi-)analytic answers often given to \textit{both} questions (i) and (ii) may need to be revised dramatically. 

In this paper we focus on the test-particle, quadrupolar ZLK problem (and on systems capable of reaching very high eccentricity, which requires $\vert j_z \vert \ll 1$). We modify the ZLK problem in exactly two ways compared to the simplest baseline just described. First, we include the physical effect of apsidal precession due to SRFs. Secondly, we relax the mathematical assumption of double averaging, instead working with the SA equations.  
On their own, the role of each of these modifications is well-understood, as we now overview. 

\subsection{Short range forces (SRFs)}

When one expands the two-body problem beyond the Newtonian point-particle limit, the leading-order conservative corrections are due to SRFs like general relativity, equilibrium tidal bulges, and rotational bulges. Each SRF manifests as apsidal precession of the argument of pericenter $\omega$, with a rate that is steeply peaked at high eccentricity \citep{fabrycky2007,liu2015,hamilton2021,grishin2018,anderson2017,blaes2002,veras2010,naoz2013pn}:
\begin{equation}
    \vert \dot{\omega}_\mathrm{SRF} \vert  \propto \frac{1}{(1-e^2)^n},
    \label{eqn:SRF}
\end{equation}
where $n=1$ for GR, $n=5$ for equilibrium tides, and $n=2$ for rotational bulges.

One can include SRFs in the inner orbit dynamics in ZLK calculations.
In the DA approximation, the dynamics are again integrable.  The two constants of motion are now $j_z$, as in the baseline case described above, and $D$, \rev{the SRF-modified counterpart of the Kozai constant,} which differs from the baseline $C_\mathrm{K}$ only at extremely high eccentricity (see \S\ref{sec:DA}). The upshot is that
SRFs always lower the maximum eccentricity $e_\mathrm{max}$ relative to its no-SRF DA value \citep{fabrycky2007} (answering question (i) above) and they leave the secular period $P_\mathrm{K}$ unchanged (answering question (ii)).

These `DA+SRF' dynamics were distilled into a simple recipe by \citet{liu2015}. Their paper has become a standard reference in population-syntheses studies of secularly driven hot Jupiters, compact-object mergers, and tidal-disruption channels (e.g., \citealt{anderson2016,petrovich2015,stephan2016,munoz2016,owen2018,teyssandier2019,vick2019,stephan2021,oconnor2021,melchor2024}).

\subsection{Single-averaging (SA)}

\begin{figure*}
\centering
\includegraphics[width=0.9\textwidth]{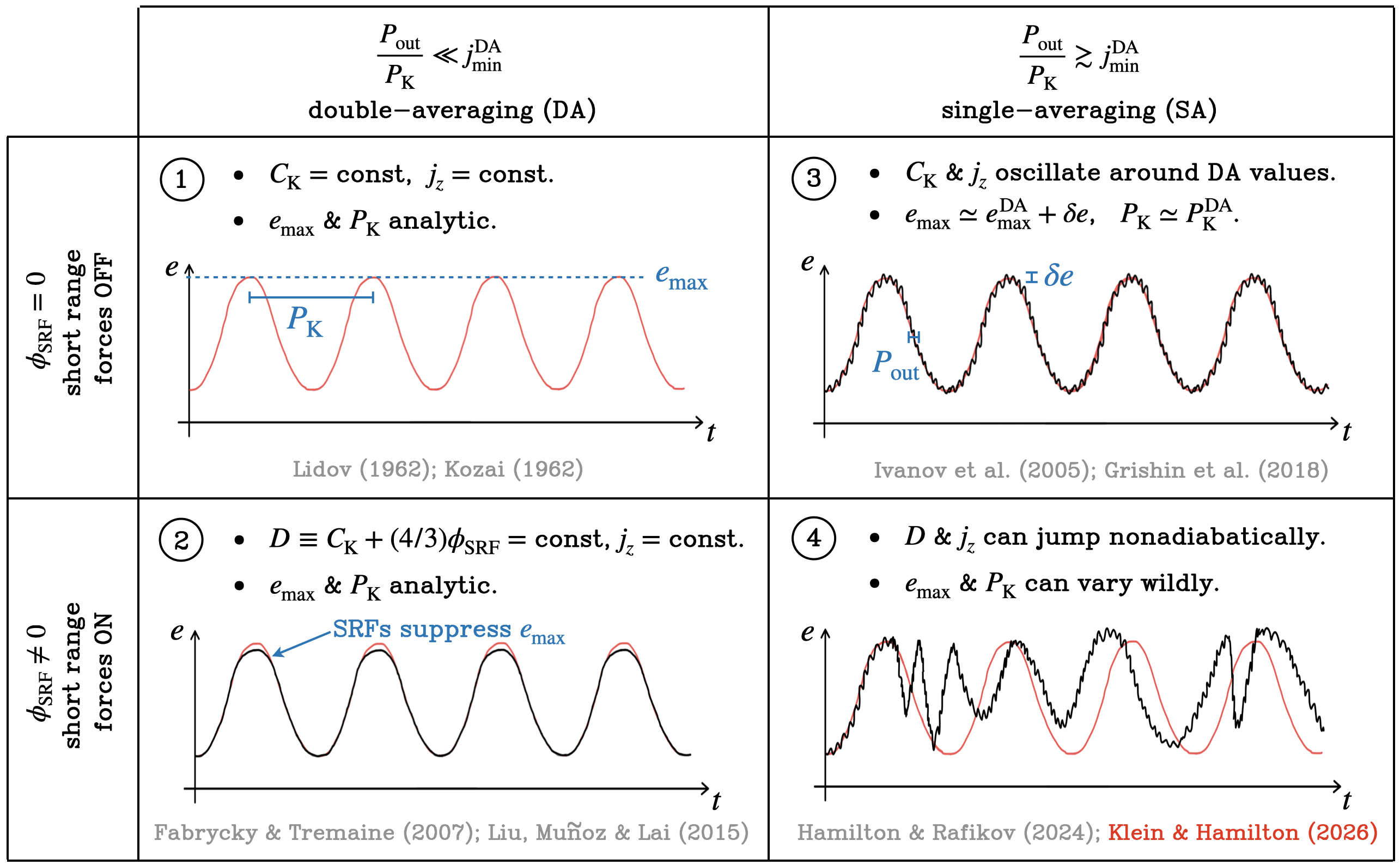}
\caption{Four dynamical regimes of the test-particle quadrupole ZLK problem (assuming $\vert j_z \vert \ll 1$). The columns separate the averaging level used (DA on the left, SA on the right) and the rows separate the physics included (no SRFs on top, SRFs on the bottom). Each panel sketches the inner binary eccentricity $e(t)$ over several secular cycles. Quadrants 1, 2, and 3 represent integrable or near-integrable problems that are well-understood (semi-)analytically. The focus of this paper is the fourth quadrant --- SA + SRFs --- wherein adiabatic invariance is completely broken.}
\label{fig:quadrants}
\end{figure*}

Parallel to this, many authors have also recognized that the DA approximation itself can fail (we will discuss the criterion for this in \S\ref{sec:naive}). In these circumstances, the natural next step is to use the single-averaging (SA) approximation, which averages only over the inner orbit. 

In an SA calculation, the torque due to the tertiary is equal to the DA expectation plus `SA fluctuations' which occur on the timescale of the outer orbital period $P_\mathrm{out}$. These in turn produce (typically very small) fluctuations in $C_\mathrm{K}$ and $j_z$ (which are now near-adiabatic invariants, rather than true integrals of motion) and hence in the binary orbital elements.  As a result, the answer to question (i) changes: the SA-corrected eccentricity maximum is approximately 
$e_\mathrm{max} \simeq e_\mathrm{max}^\mathrm{DA} + \delta e$, where $e_\mathrm{max}^\mathrm{DA}$ is the DA prediction and $\delta e$ is the SA fluctuation amplitude \citep{ivanov2005,katz2012,antonini2014,haim2018,grishin2018,hamilton2019a}.  The correction $\delta e$ can be of great importance at high eccentricity if it is comparable to or greater than $1-e_\mathrm{max}^\mathrm{DA}$, which is exactly the regime in which many GW-merger and hot-Jupiter calculations reside. Again, though, the secular period $P_\mathrm{K}$ (question (ii)) does not change.

Another, more slowly-acting SA effect is that the torque fluctuations can accumulate over many secular periods, producing a slow modulation of the ZLK cycles. This modulation can be absorbed into a corrected-DA potential, the so-called Brown Hamiltonian \citep{brown1936a,brown1936b,brown1936c,soderhjelm1975,cuk2004,breiter2015,luo2016,will2021,tremaine2023,klein2024b,grishin2024a,lei2025a,lei2025b,gao2025}. However, 
this phenomenon is not relevant for the current paper. We are going to describe a distinct effect that happens on a much shorter timescale.

\subsection{This paper: SA + SRFs}

The basic idea of this paper is summarized diagrammatically in Figure \ref{fig:quadrants}.
The four quadrants organize the test-particle quadrupole ZLK problem by what averaging is used (DA on the left, SA on the right) and what physics is included (no SRFs on top, SRFs on the bottom). As just discussed, quadrants 1, 2, and 3 are well-understood territory.
Quadrant 1 (DA, no SRFs) is the original, integrable ZLK problem; quadrant 2 (DA + SRFs) is integrable but with the high-$e$ excursions suppressed by apsidal precession; and quadrant 3 (SA, no SRFs) is nearly-integrable, departing from the DA baseline only through small SA fluctuations $\delta e$ on the timescale $P_\mathrm{out}$. This paper is concerned with the fourth quadrant (SA + SRFs).  One might expect the dynamics here to be a straightforward superposition of quadrants 2 and 3: small SA fluctuations sitting on top of an SRF-suppressed DA curve. But in general this is not what happens.  Instead, SA fluctuations and SRF precession interact non-trivially, driving large discrete jumps in $C_\mathrm{K}$ and $j_z$, and producing dynamics qualitatively different from anything in the other three quadrants.

The mechanism is the following. Away from high eccentricity the SA orbital elements oscillate tightly around their `parent' DA trajectories, which can be labeled by the (near-)adiabatic invariants $j_z$ and $C_\mathrm{K}$.
But during a brief excursion in which the inner binary reaches $e\to 1$, the SRF precession rate becomes extremely large (Eq.~(\ref{eqn:SRF})). If the apsidal angle $\omega$ changes by an order-unity amount on the outer-orbit timescale, the DA approximation --- which presumes $\omega$ varies only on the much longer secular timescale --- fails. The SA dynamics then does not oscillate back around its original parent DA trajectory after the excursion; instead, it begins to oscillate around a \textit{different} DA trajectory (labeled by a different pair $(C_\mathrm{K}, j_z)$). Thus the system has `jumped' from one DA cycle to another in a single high-$e$ episode, and this new cycle may have a totally different minimum/maximum eccentricity and secular period to the old one. As a result, 
the answers to questions (i) and (ii) above vary substantially from one secular cycle to the next.

These nonadiabatic jumps were first reported by \citet{hamilton2024} (hereafter \citetalias{hamilton2024}), who studied the test-particle quadrupole ZLK problem with GR precession. They were also identified in the simulations of \citet{rasskazov2024orbital}.
These papers called the effect \textit{relativistic phase space diffusion}, but we now think this was a poor choice of name.
The jumps are neither diffusive (a single jump can shift $C_\mathrm{K}$ by $\mathcal{O}(1)$ in one secular cycle) nor necessarily driven by GR, since any SRF whose apsidal precession rate diverges at high eccentricity through Eq.~(\ref{eqn:SRF}), equilibrium tides and rotational bulges included, will drive jumps of the same kind.\footnote{This conservative, single-averaging effect is distinct from the relativistic ``precession resonance'' of \citet{kuntz2022}, in which gravitational-wave radiation shrinks the inner orbit until its apsidal precession rate is commensurate with the outer orbit. That mechanism requires both dissipation and \rev{very} strong apsidal precession \rev{($\epsGR \gtrsim 1$, see equation \eqref{eq:epsGR})}, whereas our jumps invoke neither. See also \S\ref{sec:discussion}.} They also occur under a much wider (and more realistic) range of conditions than were appreciated by \citetalias{hamilton2024}, and they \textit{do} cause jumps in $j_z$, whereas \citetalias{hamilton2024} believed they did not. It is therefore worthwhile to revisit this problem and some of its implications in light of our new understanding.

 The rest of this paper is organized as follows.
 In \S\ref{sec:setup} we set up the dynamical framework of our problem, introduce the SA and DA equations, and discuss the validity criteria for both of them.
 In \S\ref{sec:numerical} we provide numerical examples, with various SRFs included, in which 
 nonadiabatic jump behavior drives radical changes in $e_\mathrm{max}$, $P_\mathrm{K}$ and associated quantities.
In \S\ref{sec:discussion} we discuss our results in the context of earlier literature and outline future extensions. We summarize in \S\ref{sec:summary}.


\section{Dynamical framework}
\label{sec:setup}

We consider an inner binary with component masses $m_0$ and
$m_1$ and semimajor axis $a$, perturbed by a distant mass $m_2$ on an
outer orbit with semimajor axis $a_2$ and eccentricity $e_2$, with
$a\ll a_2$.  We work in the test-particle limit so the only evolution of the outer orbit is in its mean anomaly.\footnote{\rev{By this we mean the limit in which the inner binary's orbital angular momentum is negligible compared with the outer orbit's ($J_\mathrm{in}\ll J_\mathrm{out}$), so that the outer orbit is unaffected by the inner binary's back-reaction. This does not require either inner-binary component to be massless.}}

We describe the inner orbit by the Laplace--Runge--Lenz
vector $\mathbf{e}$ and the dimensionless angular momentum vector
$\mathbf{j}=\mathbf{J}/\sqrt{G(m_0+m_1)\,a}$, satisfying
$j^2+e^2=1$ and $\mathbf{j}\cdot\mathbf{e}=0$.  
We choose the $z$-axis along the fixed outer-orbit angular-momentum
vector and the $x$-axis toward the fixed outer eccentricity vector. {The inner orbit's longitude of ascending node $\Omega$ is measured from this $x$-axis.} 
The evolution of $\mathbf{j}$ and $\mathbf{e}$ is
governed by the \citet{milankovitch1939} equations:
\begin{align}
\frac{d\mathbf{j}}{d\tau}
&= \mathbf{j}\times\nabla_{\mathbf{j}}\phi
  + \mathbf{e}\times\nabla_{\mathbf{e}}\phi,
\label{eq:milankovitch-j}\\
\frac{d\mathbf{e}}{d\tau}
&= \mathbf{j}\times\nabla_{\mathbf{e}}\phi
  + \mathbf{e}\times\nabla_{\mathbf{j}}\phi,
\label{eq:milankovitch-e}
\end{align}
where $\tau \equiv t/P_\mathrm{sec}$ is the time normalized by a characteristic secular timescale
\begin{equation}
\begin{split}
\Psec
&= \frac{1}{2\pi}\frac{m_0+m_1}{m_2}
\left(\frac{a_2}{a}\right)^3
(1-e_2^2)^{3/2}\,P_\mathrm{in} \\
&= \frac{\sqrt{G(m_0+m_1)a}}{\Phi_0},
\end{split}
\label{eq:Psec}
\end{equation}
{where $P_\mathrm{in}$ is the inner orbital period, and}
\begin{equation}
\Phi_0 \equiv \frac{Gm_2a^2}{a_2^3(1-e_2^2)^{3/2}},
\label{eq:Phi0}
\end{equation}
measures the strength of the tertiary's perturbation upon the binary, and
$\phi \equiv \Phi/\Phi_0$ is the dimensionless perturbing potential (i.e., the true perturbing potential $\Phi$ normalized by $\Phi_0$). 
In general $\phi$ includes contributions from the tertiary's tidal potential, as well as any SRF.
For this paper we expand the tidal potential in the small parameter $a/a_2 \ll 1$ and keep only the quadrupolar term.

We now make the SA approximation, i.e., we average $\phi$ over the inner orbit while fixing the outer body's instantaneous position $\mathbf{r}_2=r_2\hat{\mathbf{r}}_2$.
The full dimensionless SA perturbing potential is then 
\begin{equation}
\phi_{\rm SA}=\phi_{\rm quad}^{\rm SA}+\phi_{\rm SRF},
\label{eq:phi-SA}
\end{equation}
where
\begin{align}
\phi_{\rm quad}^{\rm SA}
&= 
\frac{3a_2^3(1-e_2^2)^{3/2}}{4r_2^3}
\left[
1 - 2e^2 - (\mathbf{j}\cdot\hat{\mathbf{r}}_2)^2
+ 5(\mathbf{e}\cdot\hat{\mathbf{r}}_2)^2
\right],
\label{eq:phi-SA-quad}
\end{align}
and 
\begin{align}
\phi_{\rm SRF}
&= \epsGR\!\left(\frac{1}{j} - \frac{1}{j_0}\right)
 + \frac{\epsRot}{3}\!\left(\frac{1}{j^3} - \frac{1}{j_0^3}\right)
\nonumber\\
&\quad + \epsTide\!\left[
   \frac{7}{24}\!\left(\frac{1}{j^9} - \frac{1}{j_0^9}\right)
   - \frac{1}{4}\!\left(\frac{1}{j^7} - \frac{1}{j_0^7}\right)
\right.
\nonumber\\
&\qquad\qquad\quad
\left.
   + \frac{1}{40}\!\left(\frac{1}{j^5} - \frac{1}{j_0^5}\right)
 \right].
\label{eq:phi-SRF}
\end{align}
\rev{Equation~\eqref{eq:phi-SA-quad} is part of the \emph{dimensionless} potential ($\phi\equiv\Phi/\Phi_0$): the outer elements $a_2$ and $e_2$ enter only through $\Phi_0$ \eqref{eq:Phi0} and cancel in the dimensional potential, which depends on the outer orbit through the instantaneous radius $r_2$ alone.}
Here the dimensionless SRF potential combines contributions from GR, rotational bulges, and tides, following \cite{fabrycky2007} and \cite{liu2015}, with $j_0 \equiv j(t=0)$ the initial value of $j$.  The terms proportional to $j_0$ have no effect on the dynamics because only the \textit{gradients} of the potential enter the equations of motion \eqref{eq:milankovitch-j}--\eqref{eq:milankovitch-e}, but we find this parameterization convenient because it means we always have $\phi_{\rm SRF}(t=0) = 0$.  The dimensionless coefficients $\varepsilon_i$ are given by
\begin{align}
\epsGR
&= 3\,\frac{m_0+m_1}{m_2}
\left(\frac{a_2}{a}\right)^3
\nonumber\\
&\quad\times
(1-e_2^2)^{3/2}
\left(\frac{G(m_0+m_1)/c^2}{a}\right),
\label{eq:epsGR}\\
\epsRot
&= k_{q,1}
\frac{m_0+m_1}{m_2}
\left(\frac{a_2}{a}\right)^3
\nonumber\\
&\quad\times
(1-e_2^2)^{3/2}
\left(\frac{\Omega_{1s}^2R_1^3}{Gm_1}\right)
\left(\frac{R_1}{a}\right)^2,
\label{eq:epsRot}\\
\epsTide
&= 15\,k_{2,1}
\frac{m_0}{m_1}
\frac{m_0+m_1}{m_2}
\left(\frac{a_2}{a}\right)^3
\nonumber\\
&\quad\times
(1-e_2^2)^{3/2}
\left(\frac{R_1}{a}\right)^5.
\label{eq:epsTide}
\end{align}
Here $c$ is the speed of light, $k_{2,1}$ and $R_1$ are the tidal Love
number and radius of the inner companion, and $k_{q,1}$ and
$\Omega_{1s}$ are its apsidal-motion constant and spin rate.
These coefficients share the same definition as in \citet{liu2015}.\footnote{\rev{Following \citet{liu2015} (and \citealt{fabrycky2007}), the rotational contribution is treated as pure apsidal precession under the assumption that the spin is aligned with the inner orbital angular momentum; we adopt that treatment unchanged. If the spin angular momentum were comparable to the orbital one the dynamics would differ qualitatively; that regime is irrelevant for the examples below. Even when that spin angular momentum is small, adiabatic averaging over a rapidly precessing spin can introduce an $\mathcal{O}(1)$ orientation-dependent factor in $\epsRot$ (e.g.,~\citealt{waisberg2025}).}} Note this means that $\varepsilon_\mathrm{GR}$ as defined here is \textit{smaller} than the $\epsilon_\mathrm{GR}$ used in \citet{hamilton2021} and \citetalias{hamilton2024} by a factor of $16$ (see Footnote~13 of \citealt{hamilton2021}).

The SA approximation itself can break down in some regimes \citep{antonini2014}. For the examples shown in this paper, however, the qualitative findings (the jumps in $C_\mathrm{K}$, the secular evolution of $j_z$, and the SRF-catalyzed extreme-eccentricity behavior) are robust to both the SA approximation and the test-particle approximation. We discuss this further (and compare with the results of direct $N$-body integration) in Appendix~\ref{sec:SA_Validity}.

\subsection{Double-averaging}
\label{sec:DA}

If we go a step further and average $\phi_\mathrm{SA}$ (equation \eqref{eq:phi-SA}) over the outer orbit, we find the DA potential
\begin{equation}
\phi_{\rm DA}=\phi_{\rm quad}^{\rm DA}+\phi_{\rm SRF},
\label{eq:phi-DA}
\end{equation}
where 
\begin{equation}
\phi^{\rm DA}_{\rm quad}
=\frac{3}{4}
\left(
\frac{1}{2}j_z^2+e^2-\frac{5}{2}e_z^2-\frac{1}{6}
\right).
\label{eq:phi-quad}
\end{equation}
The potential $\phi_\mathrm{DA}$ is axisymmetric about the $z$-axis whether or not SRFs are included, so $j_z$ is always conserved in our DA problems.

Without SRFs ($\phi_\mathrm{SRF}=0$) we have $\phi_\mathrm{DA}=\phi_{\rm quad}^{\rm DA}$. This quantity is an integral of motion, but it is more convenient to strip away the constant and the $j_z^2$ piece and work instead with the `Kozai constant'
\begin{align}
\CK &\equiv \frac{4}{3}\phi_{\rm quad}^{\rm DA}-\frac{1}{2}j_z^2+\frac{1}{6}  \nonumber \\ &= e^2-\frac{5}{2}e_z^2 = e^2\left(1-\frac{5}{2}\sin^2 i\,\sin^2\omega\right),
\label{eq:CK}
\end{align}
where $i$ is the mutual inclination and $\omega$ is the argument of pericenter.
The two integrals of motion $j_z, \CK$ dictate the entire no-SRF DA dynamics.

When SRFs \textit{are} included ($\phi_\mathrm{SRF}\neq 0$), the analogous integral of motion is
\begin{equation}
D \equiv \frac{4}{3}\phi_\mathrm{DA}-\frac{1}{2}j_z^2+\frac{1}{6} = \CK+\frac{4}{3}\phi_{\rm SRF}.
\label{eq:D-full}
\end{equation}
The integrable dynamics is then dictated entirely by $D$ and $j_z$.
Formally the range of allowed $D$ values is much larger than for $\CK$, because at very high eccentricity (low $j$), the additional term $\phi_\mathrm{SRF}$ can become extremely large (see equation \eqref{eq:phi-SRF}). However, away from peak eccentricity $\phi_\mathrm{SRF}$ is usually very small\footnote{at least for the astrophysically-relevant cases we care about, where the SRF coefficients $\varepsilon_i$ are  $\ll 1$. There are of course systems with $\varepsilon_i \gtrsim 1$ but these tend not to reach very high eccentricities anyway \citep{fabrycky2007,hamilton2021}.}, so $D\simeq \CK$ to an excellent approximation. Thus $\CK$ remains a useful label of the integrals of motion in the DA problem even with SRFs included.

In either case, the value of $\CK$ is important because (i) its sign determines the topology of the phase space motion, separating librating cycles ($\CK<0$), where $\omega$ oscillates, from rotating cycles ($\CK>0$), where $\omega$ circulates; (ii) it changes the secular period $P_\mathrm{K}$, often by a large factor near separatrices; (iii) it plays the key role in determining the \textit{minimum} eccentricity $\emin$; (iv) it modulates the maximum eccentricity $\emax$ (though the key role here is played by $j_z$).

The range of allowed $\CK$ values depends on $j_z$. The maximum possible range is achieved at $j_z=0$, namely $\CK \in [-3/2, 1]$. As $\vert j_z\vert $ is increased, this range gets narrower.  This means that a jump in $\CK$ beyond its allowed range must be associated with a \textit{decrease} in $\vert j_z\vert $.  We will see this occur explicitly in \S\ref{sec:numerical}.

\subsection{Validity of the DA approximation, and semi-analytic SA corrections to it}
\label{sec:naive}

We expect the DA approximation in \S\ref{sec:DA} to be valid only when the outer orbital period $P_\mathrm{out}$ is much shorter than the time over which DA orbital elements evolve significantly. 
The latter evolution happens most rapidly at high eccentricity, on a characteristic timescale 
$\sim (1-e_\mathrm{max}^2)^{1/2}P_\mathrm{sec}$ \citep[e.g.,][]{liu2018,hamilton2021}.
Thus, formally, we expect DA theory to be valid only if
\begin{equation}
\frac{\Pout}{\Psec}\ll j_{\rm min}^{\rm DA},
\label{eq:da-criterion}
\end{equation}
where $j_\mathrm{min}^\mathrm{DA}\equiv (1-e_{\rm max,DA}^2)^{1/2}$ is computed in the DA approximation (e.g., from \citealt{liu2015} when SRFs are included).
Alternatively, one might have demanded that the typical SA fluctuation in the binary's angular momentum at high eccentricity be sufficiently small, $|\delta j|\ll j_{\rm min}^{\rm DA}$. However, one can show that $|\delta j|\sim \Pout/\Psec$ (\citealt{ivanov2005,grishin2018}; \citetalias{hamilton2024}), so this requirement is equivalent to \eqref{eq:da-criterion}.\footnote{Strictly we should use the Kozai period $P_\mathrm{K}$ in the criterion \eqref{eq:da-criterion} rather than the characteristic timescale $P_\mathrm{sec}$. However, we nearly always have $P_\mathrm{K} \sim P_\mathrm{sec}$, and the latter is much easier to evaluate quickly. The exception is near the separatrix $\CK=0$ where we can have $P_\mathrm{K} \gg P_\mathrm{sec}$.  The main thing to keep in mind is that criteria like \eqref{eq:da-criterion} are only heuristic guides and should not be trusted blindly.}

When the heuristic criterion \eqref{eq:da-criterion} is broken, i.e., when 
\begin{equation}
\frac{\Pout}{\Psec}\gtrsim j_{\rm min}^{\rm DA},
\label{eq:da-criterion-broken}
\end{equation}
one might still hope to salvage some part of the DA program \citep{ivanov2005,grishin2018,hamilton2019a}. The hope is that $j_z$ and $\CK$ remain approximate adiabatic invariants, exhibiting only small oscillations about their DA values, so that the secular period is essentially unchanged and SA fluctuations only affect the peak eccentricity, which can be calculated (semi-)analytically. Thus we might expect
\begin{align}
    P_\mathrm{K} &\simeq P_\mathrm{K}^\mathrm{DA},
    \label{eqn:bad_period_approx}
    \\
    e_\mathrm{max} &\simeq e_\mathrm{max}^\mathrm{DA} + \delta e,
    \label{eqn:bad_emax_approx}
\end{align}
where $\delta e$ is the SA eccentricity fluctuation corresponding to $\delta j$. 
In this naive semi-analytic picture, the only role of SRFs is to reduce $e_\mathrm{max}^\mathrm{DA}$.

Our central technical claim is that, for systems lying in the fourth quadrant of Figure \ref{fig:quadrants}, this semi-analytic approach can deliver results that are highly misleading.  The underlying reason is the one already identified in \S\ref{sec:DA}: nonadiabatic jumps can place these systems on entirely different DA cycles, with a generally different $e_\mathrm{min}$, $e_\mathrm{max}^\mathrm{DA}$, $P_\mathrm{K}$, etc.
Both naive corrections \eqref{eqn:bad_period_approx} and \eqref{eqn:bad_emax_approx} therefore fail. Related to this, we believe the criterion \eqref{eq:da-criterion} has been interpreted much too loosely in the past, with double averaging assumed safe when $P_\mathrm{out}/P_\mathrm{sec}$ is only somewhat smaller, but not \textit{much much} smaller, than $j_\mathrm{min}^\mathrm{DA}$.  We warn against all such shortcuts.


\section{Numerical results}
\label{sec:numerical}

In this section we present numerical examples of nonadiabatic jump behavior and the 
cycle-to-cycle changes in $e_\mathrm{min},e_\mathrm{max}, P_\mathrm{K}$ etc. that they cause.
We show three examples, isolating one SRF at a time. For each example we integrate equations~\eqref{eq:milankovitch-j} and \eqref{eq:milankovitch-e} up to $\tau = t/\Psec = 400$. We perform integrations employing either the DA potential \eqref{eq:phi-DA} or the SA potential \eqref{eq:phi-SA}, and in each case we run both with and without SRFs.

Each example below uses astrophysically realistic masses, semimajor axes, and eccentricities.\footnote{\rev{For the planetary examples (\S\S\ref{sec:tides-example}--\ref{sec:rotation-example}), the small initial $j_z$ is chosen to demonstrate the jump mechanism. The corresponding $j_{\rm min}^{\rm DA}$ already implies star-grazing or collision at the DA level; SA+SRF then catalyzes peaks that were already in that regime. We omit dissipative and collisional physics because our focus is the conservative secular dynamics.}} The dynamics, however, depend only on the nine dimensionless numbers collected in Eq.~\eqref{eq:dimensionless-set}, built on the dimensionless framework of \S\ref{sec:setup}, so these astrophysical choices can be rescaled to apply to other systems (see Appendix~\ref{sec:specifying_ICs} for details). In our code, we specify each system directly using the dimensionless set \eqref{eq:dimensionless-set}, reconstruct the initial $\mathbf{j}$ and $\mathbf{e}$ from those, and integrate in this dimensionless space. Thus, when we report below the initial conditions in terms of orbital elements, they are really just consequences of our choice of the numbers \eqref{eq:dimensionless-set}. For reproducibility, the exact initial conditions are reported in Appendix \ref{sec:dimensional-realizations} and Table \ref{tab:dimensional}.

\subsection{Example 1: GR precession}
\label{sec:gr-example}

\begin{figure*}
\centering
\includegraphics[width=.85\textwidth]{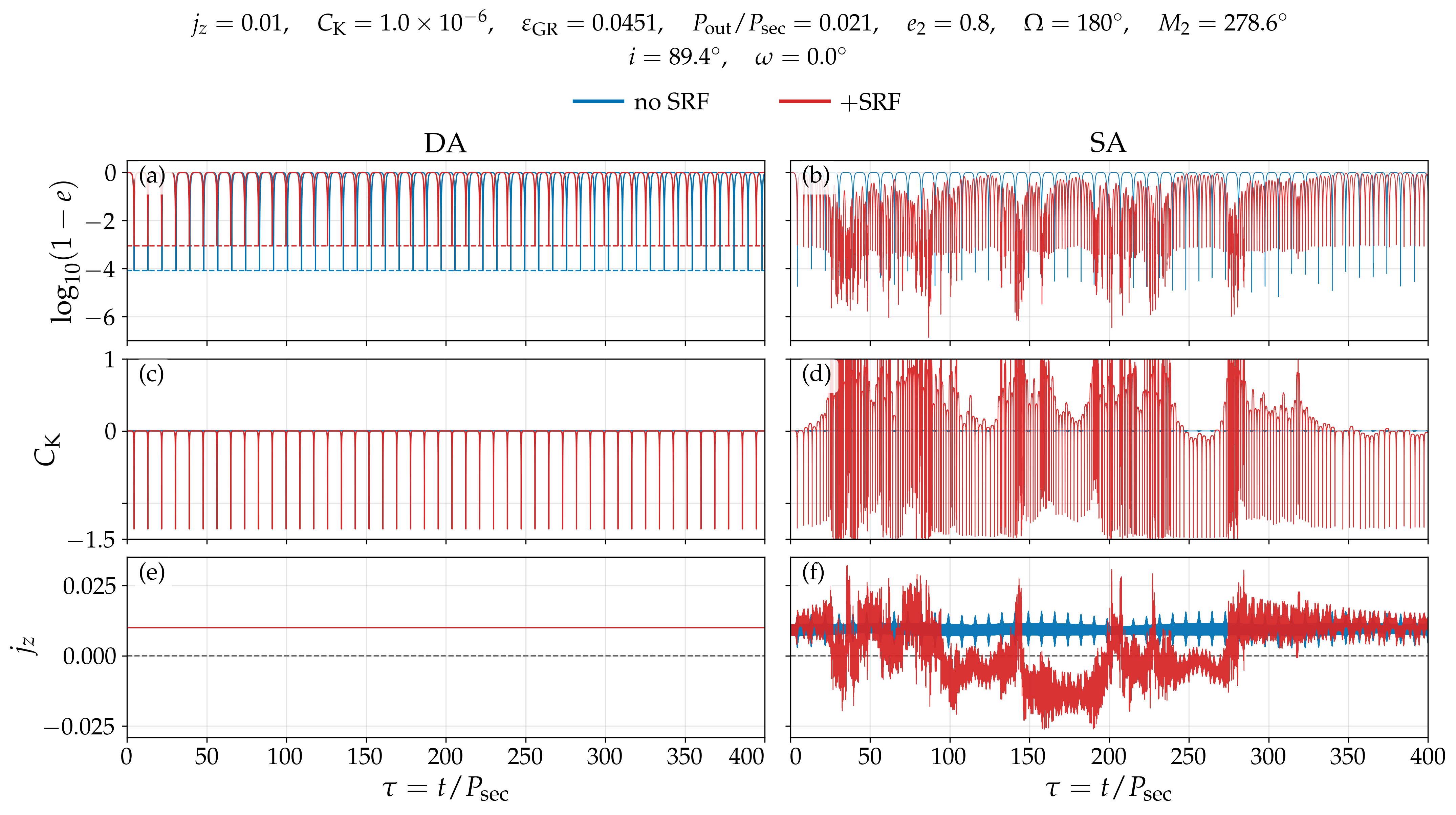}
\caption{Time evolution of $\log_{10}(1-e)$, $\CK$  and $j_z$ for the BBH--SMBH triple of \S\ref{sec:gr-example}, for initial outer mean anomaly $M_2 = 278.6^\circ$.  
GR apsidal precession is the only SRF. The left (right) column shows DA (SA) results. Red (blue) curves show results from runs with GR precession switched on (off). The two dashed horizontal lines in panel~(a) mark the DA-predicted eccentricity maximum $1-e_\mathrm{max}^\mathrm{DA}$. The dashed black lines in panels~(e) and (f) mark $j_z = 0$.}
\label{fig:gr-evolution}
\end{figure*}
\begin{figure*}
\centering
\includegraphics[width=0.8\textwidth]{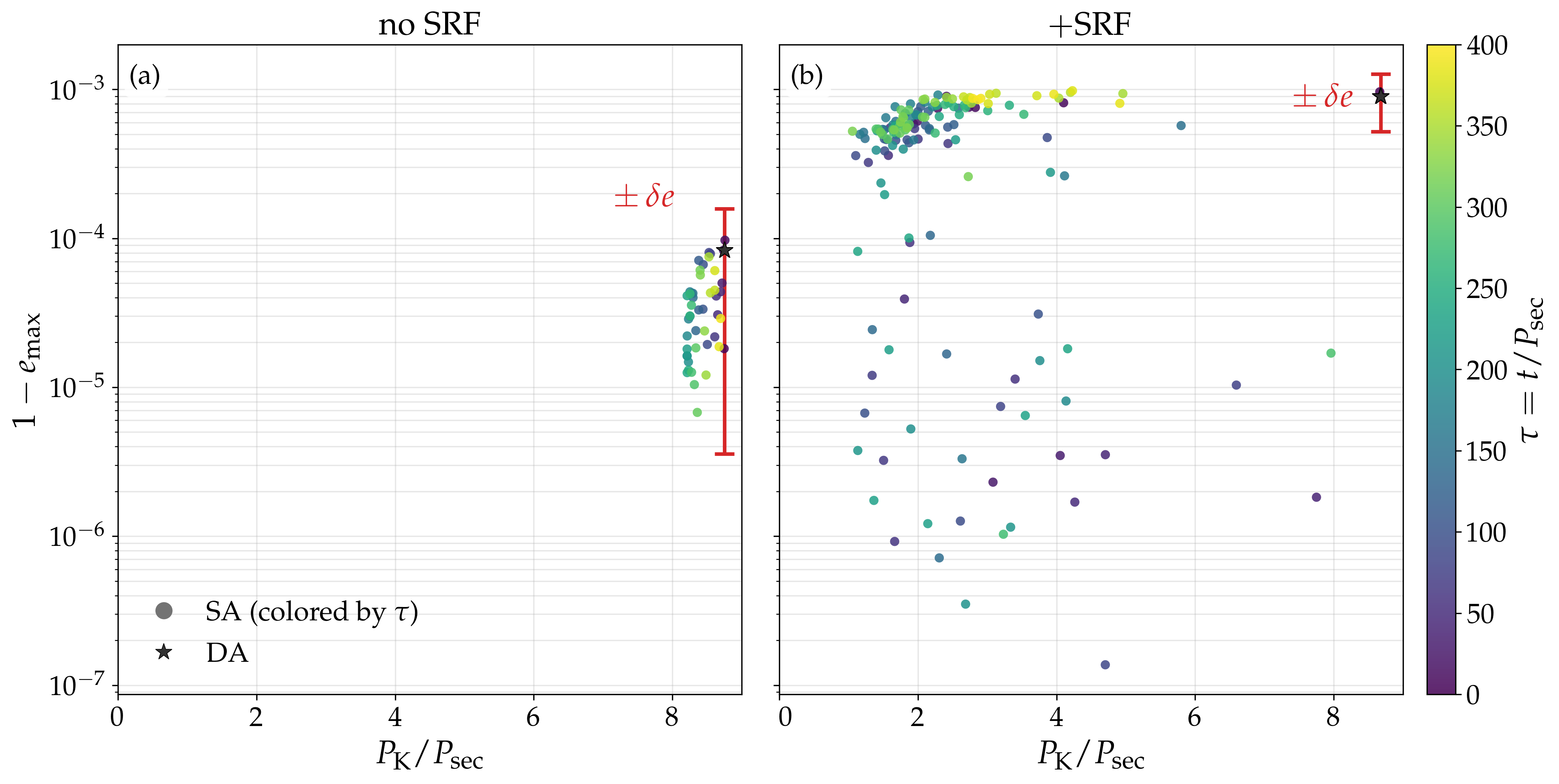}
\caption{Scatter plot of the eccentricity maximum, plotted as $1 - e_\mathrm{max}$, against the secular period $P_\mathrm{K}$, both measured at the end of each secular cycle. \rev{Each point represents a single secular cycle, colored by the dimensionless time $\tau = t/P_\mathrm{sec}$ at which it occurs (colorbar).} The system under study is the same as in Figure~\ref{fig:gr-evolution}.  The red `error bars' in each panel mark the semi-analytic prediction $e_\mathrm{max}^\mathrm{DA} \pm \delta e$ that is supposed to bracket the SA fluctuations in eccentricity; here $\delta e$ is obtained by inserting the $j_z$-oscillation amplitude of \citet[their Eqs.~32--33]{luo2016} for an eccentric outer orbit into the fluctuating-eccentricity prescription of \citet[their Eq.~36]{grishin2018}.}
\label{fig:gr-scatter}
\end{figure*}

We first consider a BBH with $m_0=m_1=30\,M_\odot$
orbiting a supermassive black hole (SMBH) of mass $m_2 = 10^7\,M_\odot$, with GR precession as
the only SRF.  For the outer orbit we take initial conditions {$a_2 \simeq 0.5\,{\rm pc}$}, $e_2=0.8$ (implying an outer period
{$\Pout \simeq 1.2\times10^4\,{\rm yr}$}) and an initial mean anomaly $M_2$ that we will specify below.  For the inner orbit we choose {$a \simeq 16\,\mathrm{AU}$} (implying a characteristic secular time {$\Psec \simeq 0.6\,{\rm Myr}$}), $e=0$, $\omega = 0.0^\circ$, $i = 89.4^\circ$, $\Omega = 180^\circ$.  The only SRF is GR precession (so $\epsTide=\epsRot=0$). These choices return the values
\begin{align}
j_z &= 0.010, \label{eq:gr-example-first} \\
\CK &= 1.0\times10^{-6}, \\
\Pout/\Psec &= 0.021, \label{eq:gr_period_ratio} \\
\epsGR &= {0.0451}, \label{eq:gr-example-epsGR} \\
j_{\rm min}^\mathrm{DA} &= 0.042.
\label{eq:gr-example-leaf}
\end{align}
Note that one can scale, e.g., the binary component masses and the two semimajor axes while keeping the dimensionless numbers \eqref{eq:gr-example-first}--\eqref{eq:gr-example-epsGR} fixed, so that the results we are about to show apply equally to, e.g., lower-mass BBHs and binary neutron stars (BNSs); precise parameters for three such realizations (including the $30+30\,M_\odot$ system shown here) are listed in Appendix~\ref{sec:dimensional-realizations}.

In Figure~\ref{fig:gr-evolution} we show the evolution of $\log_{10}(1-e)$, $\CK$, and $j_z$ as a function of time for this system. The panels on the left show the DA results, while those on the right are the SA results for a single initial outer mean anomaly $M_2 = 278.6^\circ$. In each panel the result of the no-SRF control experiment is shown in blue, while the result including the SRF is shown in red. 

We see from the left column in this Figure that in the DA approximation all trajectories are
regular, $j_z$ is constant, and the primary role of GR precession is to lower the maximum eccentricity reached, as expected (\citealt{fabrycky2007}; quadrant 2 of Figure \ref{fig:quadrants}). As for the SA runs (right column), the results of the non-SRF control case (blue) look similar to those in the DA runs, apart from some extra fluctuations in the maximum eccentricity and in $j_z$ around the DA predicted value (\citealt{ivanov2005}; quadrant 3 of Figure \ref{fig:quadrants}).  But the run including GR precession (red)
looks completely different: the minimum eccentricity, maximum eccentricity, and secular period all change from one cycle to the next, reflecting the fact that the `adiabatic invariants' $C_\mathrm{K}$ and $j_z$ are now not even approximately conserved. Panels (d) and (f) make clear that jumps in $\CK$ towards the edges of its allowed range coincide with \textit{decrements} in the magnitude of $j_z$ (\S\ref{sec:DA}). {Over the full run these jumps carry $\CK$ across almost the entire interval $[-3/2,1]$, so the system alternates between circulating and librating ZLK trajectories, while $j_z$ repeatedly changes sign. This behavior is entirely absent from the no-SRF and DA runs.}\footnote{Strictly, with SRFs the conserved DA invariant is $D = \CK + (4/3)\phi_\mathrm{SRF}$ (equation \eqref{eq:D-full}) rather than $\CK$ itself, so in the DA run $\CK$ undergoes short excursions during each high-$e$ episode (where $\phi_\mathrm{SRF}$ becomes large) while it is $D$ that stays fixed.}

How do these jumps affect (i) the maximum eccentricity $e_\mathrm{max}$ and (ii) the secular period $P_\mathrm{K}$? To find out, in Figure \ref{fig:gr-scatter} we scatter plot the peak eccentricity reached during each secular ZLK cycle versus the secular period of that cycle (defined as the time since the previous eccentricity peak).
The DA calculations produce single, analytic values for $e_\mathrm{max}$ and $P_\mathrm{K}$, and we again show these results with stars, while the SA results are shown with dots, colored by the time of the measurement.  Finally with the red error bars in each panel we show the naive semi-analytic prediction for the spread in eccentricity maxima, $e_\mathrm{max}^\mathrm{DA}\pm\delta e$, with {$\delta e \simeq 3.7\times10^{-4}$} the `SA fluctuation' obtained by inserting the $j_z$-oscillation amplitude of \citet[their Eqs.~32--33]{luo2016} for eccentric outer orbits (here $e_2=0.8$) into the fluctuating-eccentricity prescription of \citet[their Eq.~36]{grishin2018}.

We see from panel (a) of Figure \ref{fig:gr-scatter} that without SRFs, the SA fluctuations drive essentially no change in secular period and only a modest spread in $e_\mathrm{max}$ values, consistent with equations \eqref{eqn:bad_period_approx}-\eqref{eqn:bad_emax_approx}.
But in panel (b), with SRFs included, the story is completely different. The value of $1-e_\mathrm{max}$ in these runs varies by $\sim 4$ orders of magnitude, dwarfing the expected semi-analytic correction $\delta e$. The secular period $P_\mathrm{K}$ varies by almost an order of magnitude across cycles, contrary to the expectation that it should be roughly constant. \rev{The deepest peak here ($1-e_\mathrm{max}\approx1.4\times10^{-7}$) does not fall below the value $1-e\approx2.5\times10^{-8}$ that the single-averaged formula of \citet{huang2026} gives as an upper bound on $e_\mathrm{max}$ (over all initial conditions) for this system, so our results are consistent with that ceiling.}  

\begin{figure*}[!t]
    \centering
\includegraphics[width=0.8\textwidth]{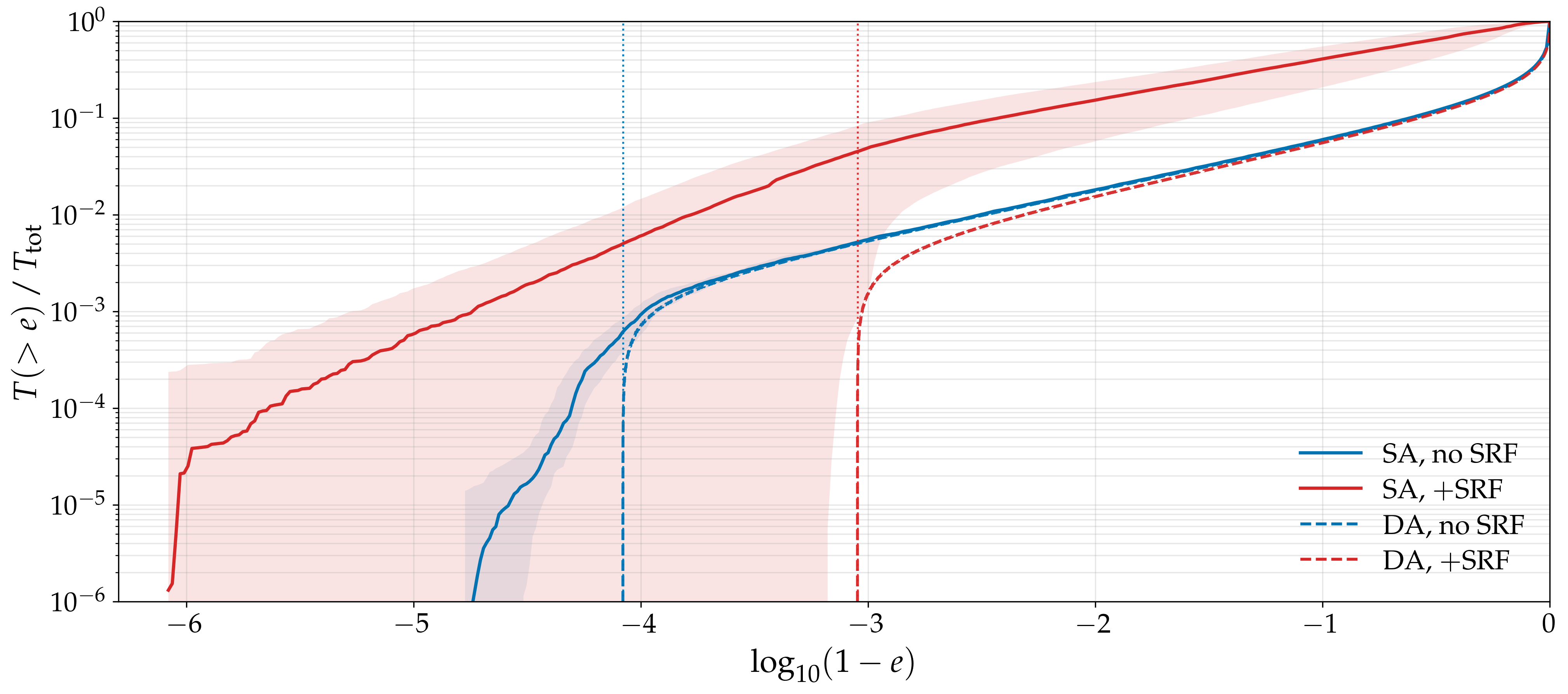}
\caption{Cumulative residence time $T(>e)$ as a fraction of the total integration time $T_\mathrm{tot}$, versus $\log_{10}(1-e)$, for the BBH--SMBH system of Figures~\ref{fig:gr-evolution}--\ref{fig:gr-scatter}, run 50 times with random initial  outer mean anomaly $M_2 \in [0,2\pi)$.
Red and blue colors correspond to SRFs (in this case GR precession) being switched on and off respectively. The solid lines show the median 
SA results and the shaded bands show their 5--95\% range.
Dashed lines show the DA results; they asymptote towards zero when they reach their predicted maximum eccentricities $e_\mathrm{max}^\mathrm{DA}$ (vertical dotted lines). }
\label{fig:gr-residence}
\end{figure*}
Figures \ref{fig:gr-evolution}--\ref{fig:gr-scatter} were for the choice of outer mean anomaly $M_2=278.6^\circ$.
Different choices of $M_2$ generally produce different evolution.  To characterize the spread of possible outcomes, we ran an ensemble of 50 realizations with $M_2$ drawn uniformly from $[0,2\pi)$ and, for each realization, computed the amount of time $T(>e)$ the inner binary spent above each eccentricity $e$.  In Figure \ref{fig:gr-residence} we plot the resulting cumulative `residence time' $T(>e)/T_\mathrm{tot}$ (where $T_\mathrm{tot} = 400 P_\mathrm{sec}$ is the total integration time) as a function of $\log_{10}(1-e)$.  The dashed curves show the DA results (which are realization-independent), the solid curves show the median of the SA ensemble, and the shaded bands show their $5$--$95\%$ ranges.
We see that the two DA results are nearly identical at all eccentricities in the range
$\log_{10}(1-e)\gtrsim-2$, which is what we expect since SRFs are irrelevant there. As we increase $e$ past this threshold (i.e., move further left in Figure \ref{fig:gr-residence}),  the red dashed curve (DA with SRFs) cuts off before the blue dashed curve (DA without SRFs) does, which again is expected since in DA theory the SRFs suppress extreme eccentricity behavior (\citealt{fabrycky2007}; \citealt{liu2015}). Importantly, though, the median SA results (solid lines) show the opposite ordering, with red sitting to the \emph{left} of blue. This is the consequence of SRFs catalyzing, rather than suppressing, extreme eccentricity behavior when SA fluctuations are included. As a result, for any eccentricity smaller than $e\simeq 0.9999$, the cumulative time spent above that eccentricity is about an order of magnitude \textit{larger} with SRFs than without!  These are two distinct effects: the leftward shift of the SA curve reflects a higher maximum eccentricity $e_\mathrm{max}$, while the upward shift --- the extra time spent above a \emph{fixed} eccentricity --- reflects not only $e_\mathrm{max}$ but also the cycle-to-cycle changes in the minimum eccentricity $e_\mathrm{min}$ and the secular period $P_\mathrm{K}$. The residence history is therefore sensitive to the whole post-jump sequence of cycles, not just the single deepest eccentricity peak.


\subsection{Example 2: Tidal precession}
\label{sec:tides-example}

\begin{figure}
\centering
\includegraphics[width=0.49\textwidth]{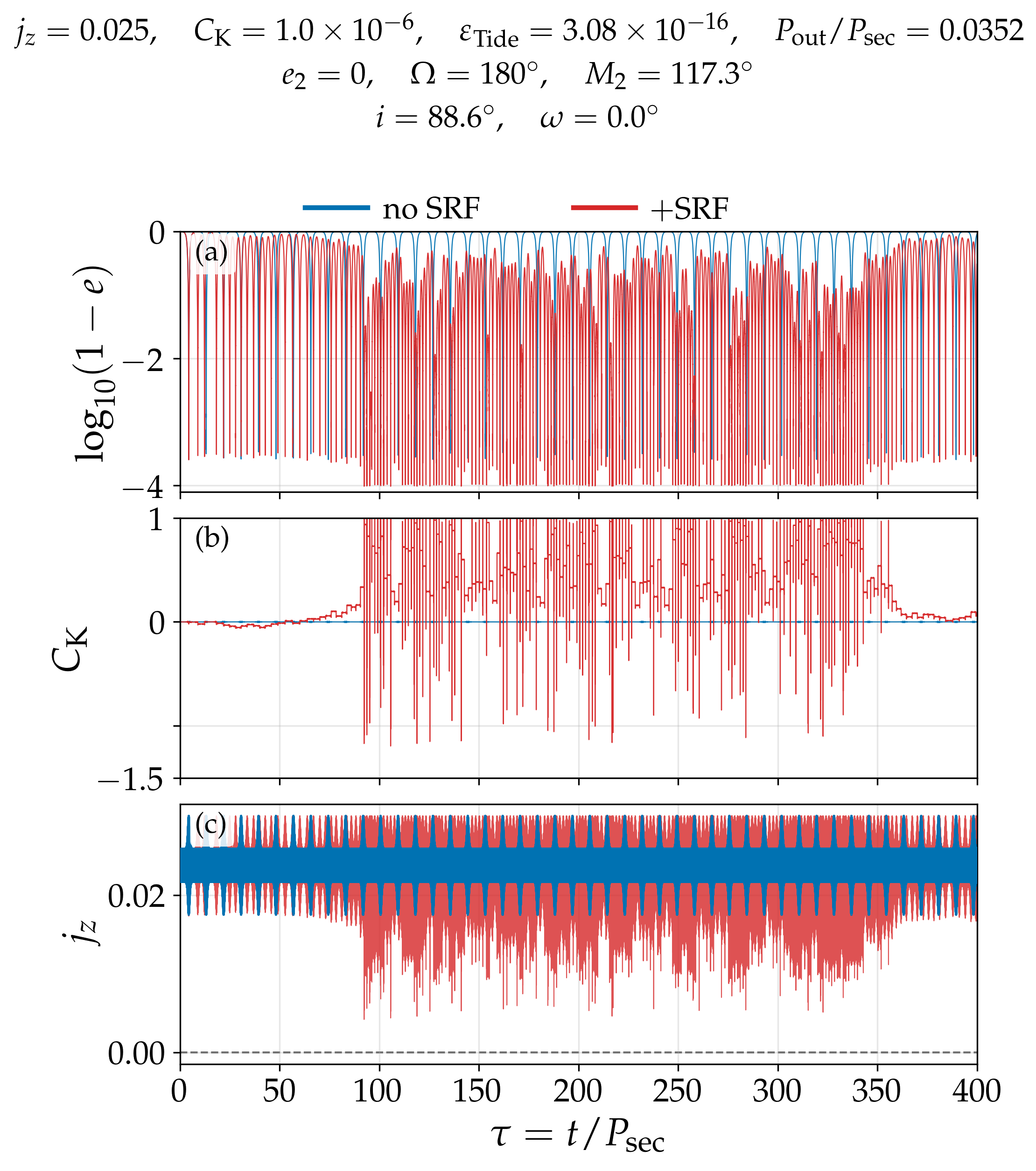}
\caption{Time evolution of $\log_{10}(1-e)$, $\CK$ and $j_z$ for the Sun--Neptune system of \S\ref{sec:tides-example}, and choosing the initial outer mean anomaly to be $M_2=117.3^\circ$.
Precession due to the tidal bulge is the only SRF.  Only the single-averaged results are shown, since here the DA results with and without SRFs are indistinguishable.  Red curves include the tide ($+$SRF), blue curves do not, and the dashed line in panel (c) marks $j_z=0$.}
\label{fig:tide-evolution}
\end{figure}
\begin{figure}
\centering
\includegraphics[width=0.49\textwidth]{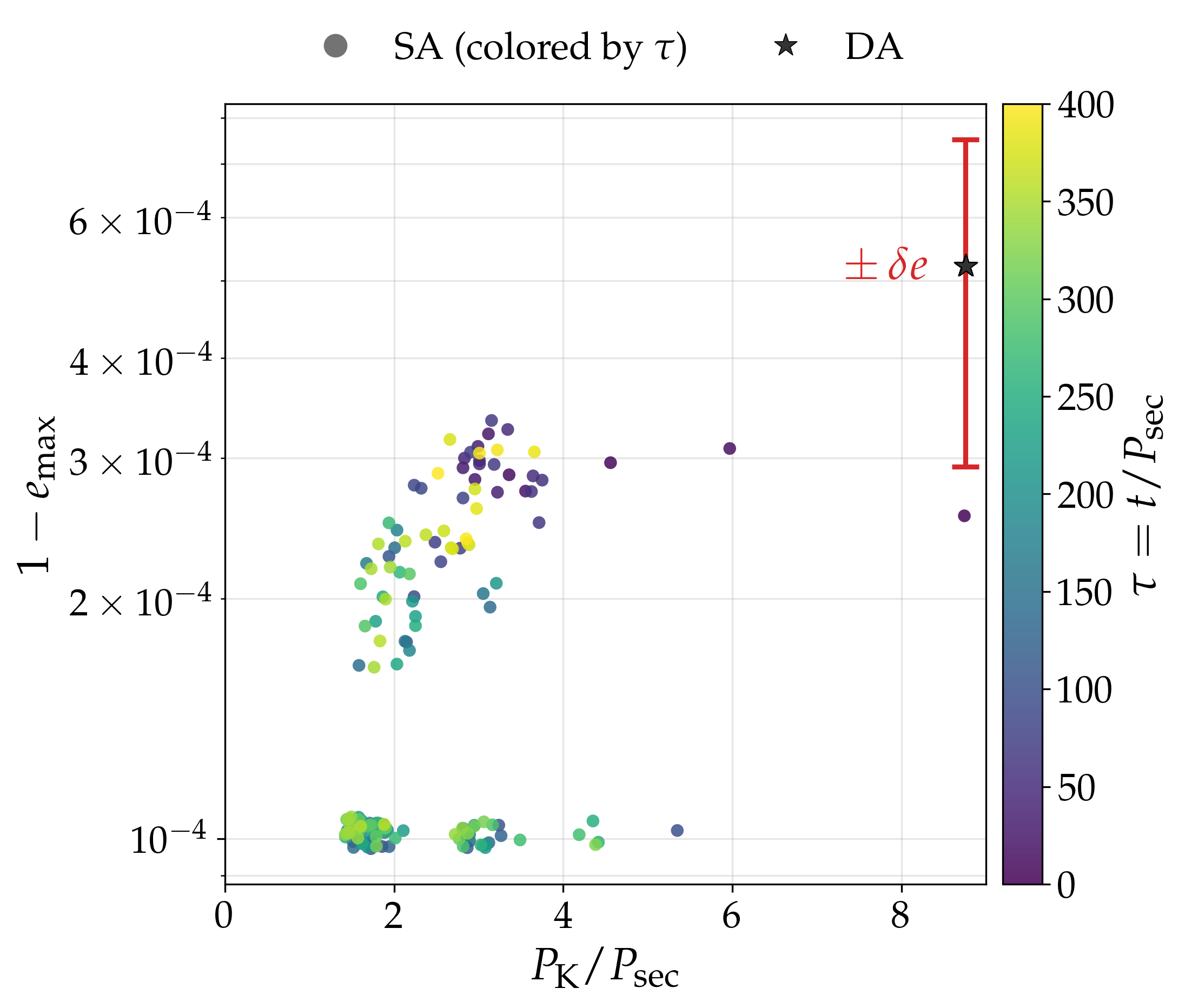}
\caption{As in panel (b) of Figure \ref{fig:gr-scatter}, except for the Sun--Neptune system of \S\ref{sec:tides-example}, for which a tidal bulge provides the only SRF.}
\label{fig:tide-scatter}
\end{figure}
\begin{figure*}
\centering
\includegraphics[width=.8\textwidth]{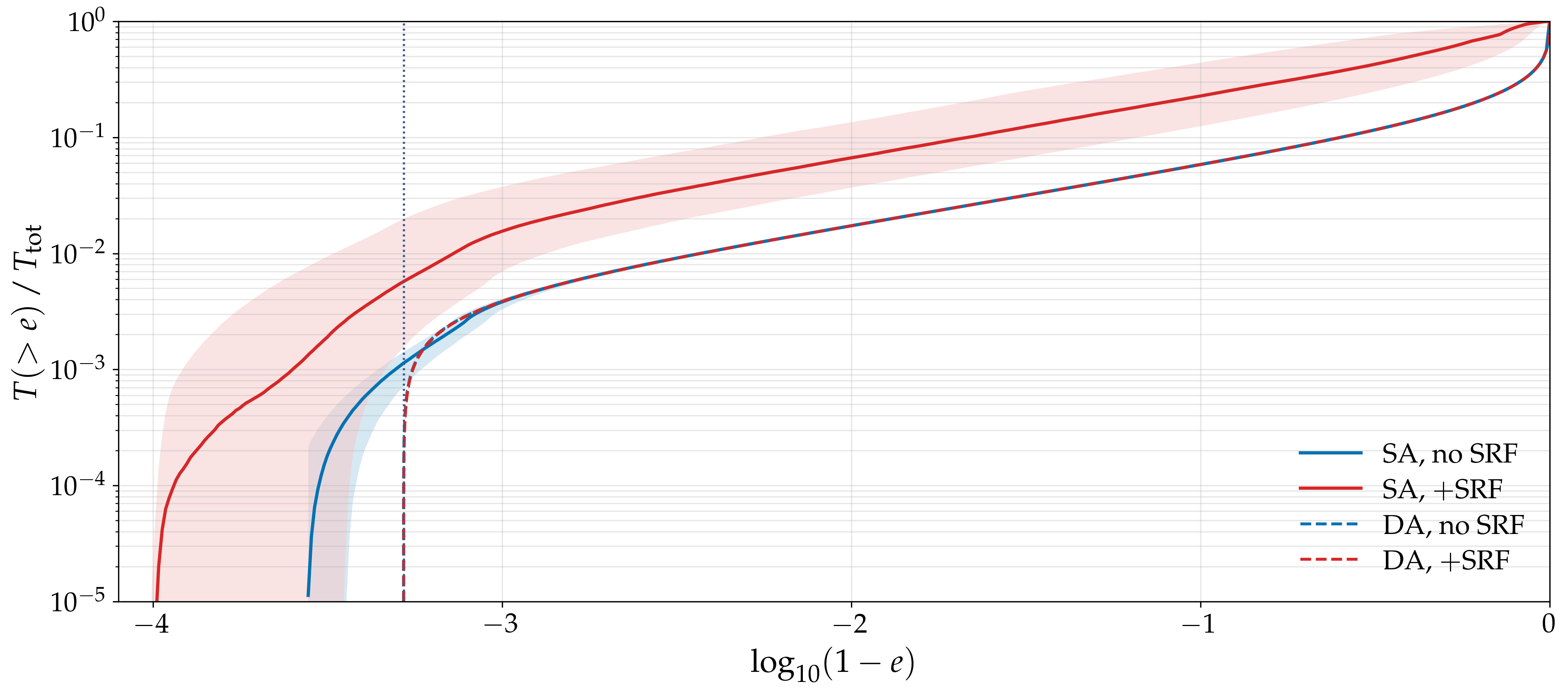}
\caption{As in Figure~\ref{fig:gr-residence}, except for the Sun--Neptune system of \S\ref{sec:tides-example}, for which a tidal bulge provides the only SRF. Unlike in Figure~\ref{fig:gr-residence} the no-SRF and $+$SRF DA curves now coincide, because at the DA level the tidal precession does nothing.}
\label{fig:tide-residence}
\end{figure*}

Next we consider a Sun--Neptune pair ($m_0=1\,M_\odot$, $m_1=m_\mathrm{Nep} = 5.1\times10^{-5}\,M_\odot$), perturbed by an $m_2=0.64\,M_\odot$ outer companion on a \emph{circular} orbit at {$a_2\simeq 240\,\mathrm{AU}$} (giving an outer period {$P_\mathrm{out}\simeq2900$ yr}).   For the inner orbit we choose {$a \simeq 12\,\mathrm{AU}$} (implying a characteristic secular time {$\Psec \simeq 8\times10^4\,{\rm yr}$}), $e=0$, $\omega = 0.0^\circ$, $i = 88.6^\circ$, $\Omega = 180^\circ$ {(precise values are in Appendix~\ref{sec:dimensional-realizations})}.
We include apsidal precession due to Neptune's tidal bulge as our only SRF (so that $\epsGR=\epsRot=0$)\footnote{Although $\epsGR$ is not strictly zero for this system (in fact $\epsGR\sim3\times10^{-5}$), including it does not change our results.}, using Love number $k_2=0.127$ \citep{gavrilov1977} and radius $R_\mathrm{Nep}\simeq2.5\times10^4\,\mathrm{km}$.  
These choices give
\begin{align}
j_z &= 0.025, \\
\CK &= 1.0\times10^{-6}, \\
\Pout/\Psec &= 0.035, \\
\epsTide &= 3.08\times10^{-16}, \\
j_{\rm min}^\mathrm{DA} &= 0.032.
\end{align}

In Figure~\ref{fig:tide-evolution} we show the SA evolution of $\log_{10}(1-e)$, $\CK$ and $j_z$ for initial outer mean anomaly $M_2=117.3^\circ$.  As in the GR example of \S\ref{sec:gr-example}, the no-SRF SA run (blue) tracks the regular DA cycle with only small fluctuations, whereas the run including the SRF (red) shows jumping behavior near eccentricity maxima. The value of $\CK$ wanders across much of its allowed range $[-3/2,1]$ and the secular period and peak eccentricity change from one cycle to the next.  The striking point is that the SRF is fantastically weak ($\epsTide=3.08\times10^{-16}$): so small, in fact, that in the DA approximation it does nothing at all. And yet, in the SA case, it drives evolution much more extreme than could have been predicted from DA theory. 

In Figure \ref{fig:tide-scatter} we show a scatter plot of maximum eccentricity versus secular period, only for the case with the SRF switched on (cf. panel (b) of Figure \ref{fig:gr-scatter}). In this case one can think of the maximum eccentricity measurements as being split into two `families': one centered around $1-e_\mathrm{max} \simeq 2.5 \times 10^{-4}$, which are \textit{reasonably} well-described by the standard semi-analytic approximation \eqref{eqn:bad_emax_approx}, and another around {$1-e_\mathrm{max} \simeq 10^{-4}$} that can only be produced by the nonadiabatic jumps. 
Meanwhile the secular period can be shorter than the DA prediction by almost an order of magnitude.

Finally, in Figure~\ref{fig:tide-residence} we plot the cumulative residence time fraction for 50 realizations of this system, each initialized with random outer mean anomaly.  The red and blue dashed DA curves coincide in this case, meaning that at the DA level, the tidal precession does nothing.  Yet the SA ensemble again behaves as it did in Figure \ref{fig:gr-residence}: the median curve with the SRF included (red) sits to the \emph{left} of and \emph{above} the no-SRF median (blue).  The same two effects operate here: the leftward shift reflects a higher $e_\mathrm{max}$, while the upward shift reflects the cycle-to-cycle changes in $e_\mathrm{min}$ and the secular period $P_\mathrm{K}$.

 This example confirms that --- just as in the GR case of \S\ref{sec:gr-example} --- apsidal precession due to tides may \emph{catalyze}, rather than suppress, extreme eccentricity evolution when SA fluctuations are included.  
 

\subsection{Example 3: Rotational bulge}
\label{sec:rotation-example}

\begin{figure}
\centering
\includegraphics[width=0.49\textwidth]{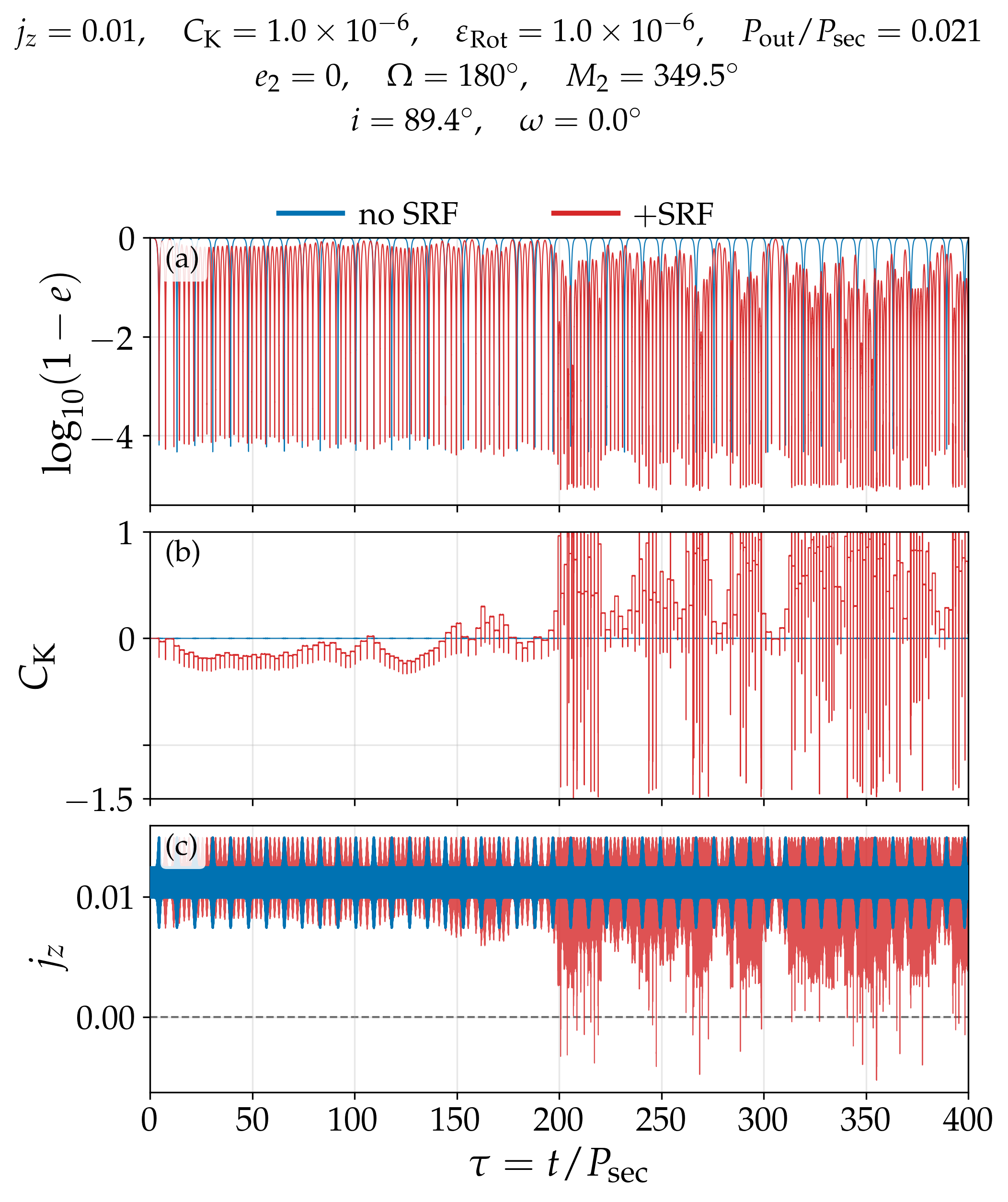}
\caption{As in Figure~\ref{fig:tide-evolution}, but for the Sun--Jupiter-like system described in \S\ref{sec:rotation-example} where  the only SRF included is due to a rotational bulge.}
\label{fig:rot-evolution}
\end{figure}
\begin{figure}
\centering
\includegraphics[width=0.49\textwidth]{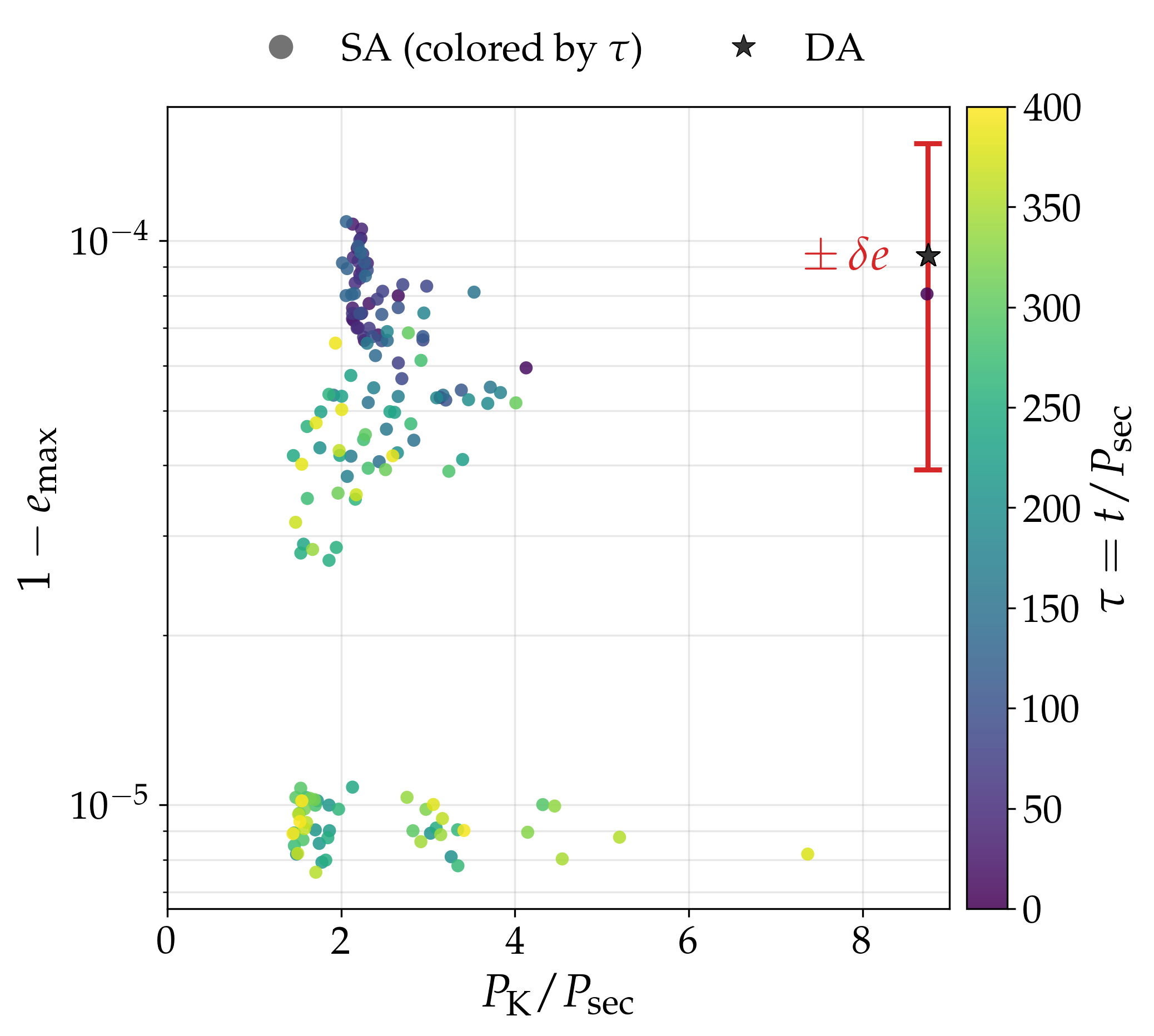}
\caption{As in Figure~\ref{fig:tide-scatter}, but for the Sun--Jupiter-like system described in \S\ref{sec:rotation-example} where  the only SRF included is due to a rotational bulge.}
\label{fig:rot-scatter}
\end{figure}

For our final example we consider a Sun-like star ($m_0=1\,M_\odot$) hosting a Jupiter-mass planet ($m_1=1\,M_J$), perturbed by an $m_2=0.20\,M_\odot$ M-dwarf on a \emph{circular} orbit at $a_2=100\,$AU (so the outer period is {$P_\mathrm{out}\simeq900$ yr}).  We choose the inner orbital elements $a=7\,$AU (so the secular time is {$P_\mathrm{sec} \simeq 4.3\times10^4\,{\rm yr}$}), {$e=0$}, $\omega = 0.0^\circ$, $i=89.4^\circ$, $\Omega=180^\circ$ {(precise values in Appendix~\ref{sec:dimensional-realizations})}. We let the planet spin near its primordial rate with period $P_\mathrm{spin}=9.93\,$hr \citep{batygin2018}, and include rotational bulges as the only SRF (so $\epsGR=\epsTide=0$).\footnote{We do this only to show that the rotational bulge \emph{alone} can drive nonadiabatic jumps. In reality the other SRFs are not negligible in this system; in particular $\epsTide\simeq10^{-13}$ turns out to be the dominant effect. This example is thus a controlled demonstration, not a fully realistic system.} Then we have 
\begin{align}
j_z &= 0.010, \\
\CK &= 1.0\times10^{-6}, \\
\Pout/\Psec &= 0.021, \\
{\epsRot} &= 1.0\times10^{-6}, \\
j_{\rm min}^\mathrm{DA} &= 0.014.
\end{align}

In Figure~\ref{fig:rot-evolution} we show the resulting SA evolution: again the no-SRF run (blue) tracks the regular DA cycle, while the run with the SRF included (red) jumps nonadiabatically, with $\CK$ wandering across much of its allowed range.

In Figure~\ref{fig:rot-scatter} we show the corresponding $e_\mathrm{max}$ versus $P_\mathrm{K}$ scatter plot, which exhibits the same catalysis as in the previous examples.
In fact, just as in the tidal example (Figure \ref{fig:tide-scatter}), there are two families of eccentricity maxima: an upper family (reaching $1-e_\mathrm{max}\sim 10^{-4}$) that could be reasonably well-described by the semi-analytic prescription $\pm \delta e$,   and a separate lower family (clustered around $1-e_\mathrm{max}\sim 10^{-5}$)  that can only be the result of nonadiabatic jumps.

\begin{figure*}
\centering
\includegraphics[width=0.8\textwidth]{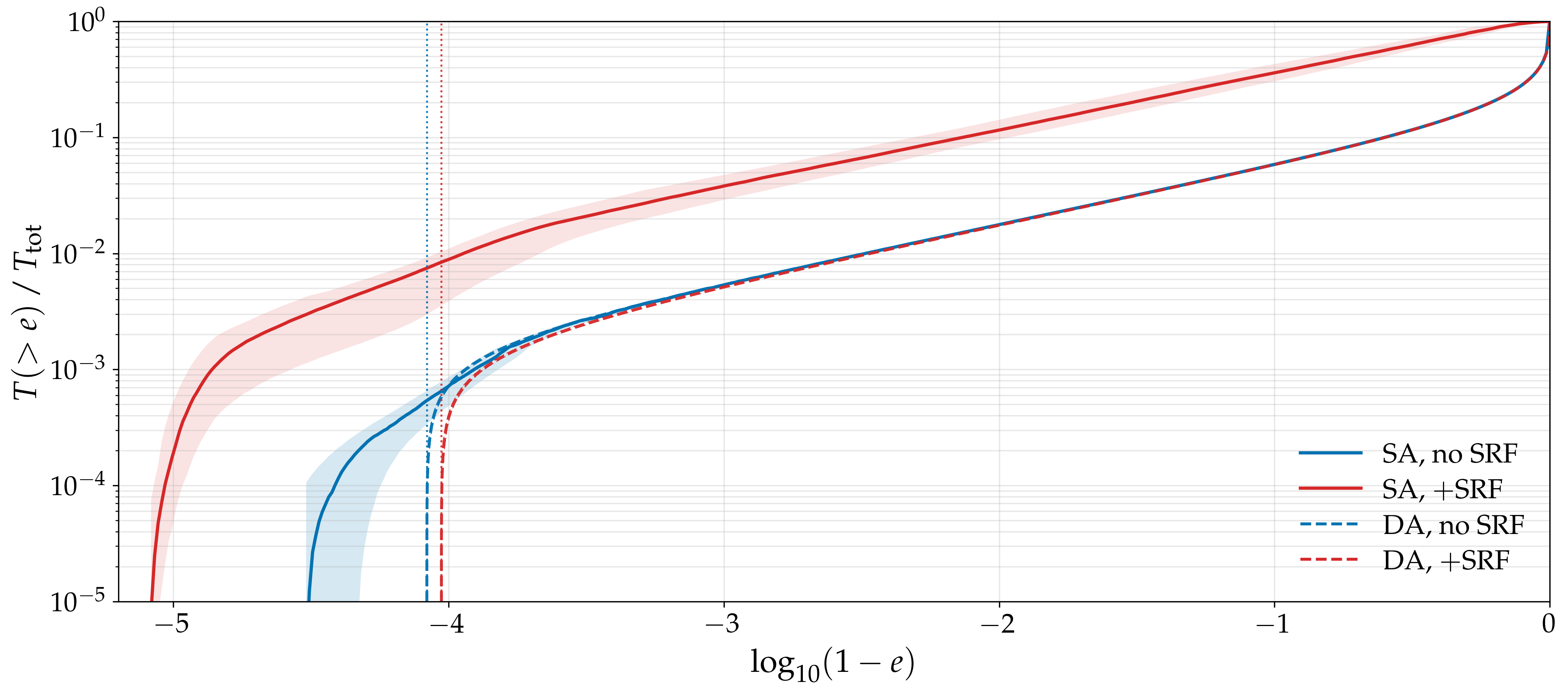}
\caption{As in Figures~\ref{fig:gr-residence} and~\ref{fig:tide-residence}, but for the example from \S\ref{sec:rotation-example} with the rotational bulge providing the only SRF.}
\label{fig:rot-residence}
\end{figure*}

Finally, in Figure~\ref{fig:rot-residence} we plot the cumulative residence-time fraction for $50$ realizations of this system with randomized initial $M_2$; as in the previous examples, the SA+SRF median result (red) sits to the \emph{left} of and \emph{above} the no-SRF median (blue), confirming that the rotational bulge too catalyzes extreme eccentricity evolution.

\subsection{The onset of nonadiabatic jumps}
\label{sec:onset}

\rev{The three examples above each sit at a single value of $\Pout/\Psec$. Our central claim is that the onset of jump behavior occurs when this ratio becomes comparable to or exceeds the DA minimum angular momentum $j_\mathrm{min}^\mathrm{DA}(j_z,\CK,\varepsilon_i)$. To demonstrate this explicitly, we ran simulations very similar to those discussed in \S\S\ref{sec:gr-example}--\ref{sec:rotation-example}, using the same dimensionless parameter sets \eqref{eq:dimensionless-set} for our initial conditions, \textit{except} we varied the ratio $\Pout/\Psec$ over a few orders of magnitude.
For each value of $\Pout/\Psec$ we ran an ensemble of 50 simulations with random initial outer mean anomaly $M_2$. We then recorded the maximal per-cycle jump in the Kozai constant,}
\begin{equation}
\max|\Delta\CK| \;\equiv\; \max_{M_2}\ \max_{k}\ \bigl|\CK^{(k+1)}-\CK^{(k)}\bigr|,
\label{eq:maxdck}
\end{equation}
\rev{where $k$ indexes secular cycles.}

\begin{figure*}
\centering
\includegraphics[width=0.75\textwidth]{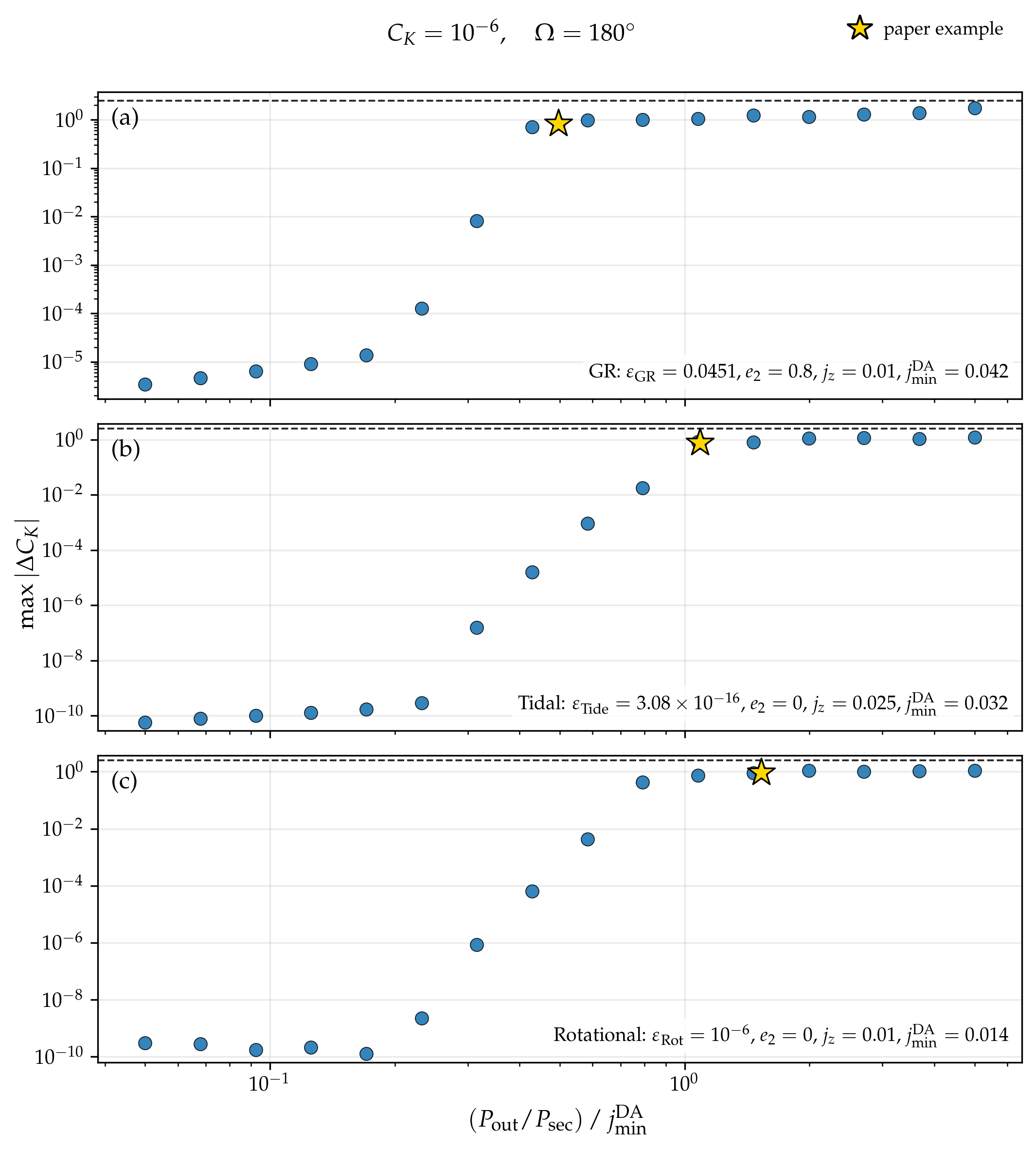}
\caption{\rev{Maximum single-cycle jump in the Kozai constant, $\max|\Delta\CK|$ \eqref{eq:maxdck}, versus $(\Pout/\Psec)/j_\mathrm{min}^\mathrm{DA}$, for the three worked examples: (a)~GR precession (\S\ref{sec:gr-example}), (b)~tidal bulge (\S\ref{sec:tides-example}), (c)~rotational bulge (\S\ref{sec:rotation-example}). In each panel the dimensionless set \eqref{eq:dimensionless-set} is held fixed at the values quoted in the box and only $\Pout/\Psec$ is varied; each point represents the maximum jump for an ensemble of 50 simulations in which we vary the initial outer mean anomaly $M_2$. Stars mark the realizations we studied in detail above  (Figures~\ref{fig:gr-evolution}, \ref{fig:tide-evolution} and~\ref{fig:rot-evolution}); the dashed line is the maximum possible $|\Delta\CK|=5/2$.}}
\label{fig:onset}
\end{figure*}

\rev{In Figure \ref{fig:onset} we show the result of these calculations as a function of the ratio $(\Pout/\Psec)/j_\mathrm{min}^\mathrm{DA}$, which measures how deeply a system violates the adiabatic condition \eqref{eq:da-criterion}.  Since $j_z$, $\CK$ and the SRF strengths $\varepsilon_i$ are held fixed within each example, the value of $j_\mathrm{min}^\mathrm{DA}$ is fixed in each panel, so the only quantity being varied is $\Pout/\Psec$.
We see that when the ratio $(\Pout/\Psec)/j_\mathrm{min}^\mathrm{DA}$ is small the jumps are absent: $\CK$ executes only tiny, bounded SA oscillations, so $\max|\Delta\CK|$ lies many decades below unity and the naive semi-analytic picture \eqref{eqn:bad_period_approx}--\eqref{eqn:bad_emax_approx} holds. As the ratio approaches unity the nonadiabatic jumps switch on, and once it exceeds unity they saturate near the maximum possible value $|\Delta\CK|\le5/2$ set by the allowed range $\CK\in[-3/2,1]$. }

\rev{The results from the three worked examples from \S\S\ref{sec:gr-example}--\ref{sec:rotation-example} are shown with stars in Figure \ref{fig:onset}. They each sit fairly close to the onset of nonadiabatic jumps, but are representative examples of an overall monotonic trend; they are \textit{not} hand-picked special cases. This figure thus makes the regime of validity explicit: double-averaging is trustworthy only when $\Pout/\Psec\ll j_\mathrm{min}^\mathrm{DA}$ (criterion \eqref{eq:da-criterion}), while the nonadiabatic jumps that are the subject of this paper are the generic outcome once $\Pout/\Psec\gtrsim j_\mathrm{min}^\mathrm{DA}$.}

\rev{We stress again that the criterion \eqref{eq:da-criterion} is approximate, and great care must be taken even when the ratio $(\Pout/\Psec)/j_\mathrm{min}^\mathrm{DA}$ is ostensibly small. 
For instance, we have found other realistic systems in which order-unity jumps in $C_\mathrm{K}$ can appear even at $(\Pout/\Psec)/j_\mathrm{min}^\mathrm{DA} \simeq 0.1$. We show this explicitly in Appendix~\ref{sec:onset-belowthreshold}.}

\section{Discussion}
\label{sec:discussion}

Hierarchical triples are remarkably rich dynamical systems whose behavior is still not fully understood. Here we have uncovered yet further layers of this richness. Namely, as the tertiary orbits in a hierarchical triple, it exerts a torque on the inner binary that fluctuates on the outer orbital timescale, and when these fluctuations act in concert with SRF-induced apsidal precession (i.e., when the system lies in the fourth quadrant of Figure~\ref{fig:quadrants}), the inner binary can suffer nonadiabatic jumps at high eccentricity. These jumps occur because the rapid apsidal precession when $e\to1$ swings the argument of pericenter $\omega$ through an $\mathcal{O}(1)$ angle on the outer orbital timescale. Then the system has no time to oscillate around its original secular (DA) trajectory, and instead emerges near a new secular trajectory, pinned to a new set of approximate adiabatic invariants $\CK$, $j_z$.
This means that the inner binaries of hierarchical triples may explore phase space far more efficiently than previously recognized.

Nonadiabatic jumps were already reported by \citetalias{hamilton2024} (see also \citealt{rasskazov2024orbital}) for one specific SRF, namely 1pN GR apsidal precession, and they called the effect `relativistic phase space diffusion'.
Our findings extend/differ from those of \citetalias{hamilton2024} as follows:
\begin{itemize}
    \item \citetalias{hamilton2024} focused on GR, but we have shown that GR is not the only SRF that can drive nonadiabatic jumps: any apsidal precession effect that is sufficiently strong at high eccentricity (see equation \eqref{eqn:SRF}) can do the same.

    \item \citetalias{hamilton2024} referred to the effect of jumps as `diffusion', implying that individual jumps are in some sense small. However, we have found that a single jump can change $C_\mathrm{K}$ by $\mathcal{O}(1)$. This is not small (recall the maximum dynamic range is $\CK \in [-3/2,1]$), so the word `diffusion' may be misleading.

    \item \citetalias{hamilton2024} believed that $j_z$ did not undergo nonadiabatic jump behavior. We have shown that this is not true. In fact, as we discuss further below, jumps in $j_z$ are both inevitable and dynamically important.

    \item \citetalias{hamilton2024} showed that nonadiabatic jumps arise for binaries subject to a range of external potentials. We specialized here to a genuine hierarchical triple (in which the outer tertiary is a point mass, i.e., a Keplerian potential), but we do not anticipate that anything would change qualitatively if we were to relax this.

\item \citetalias{hamilton2024} focused on quantifying the jumps in the space of approximate adiabatic invariants over a single eccentricity peak. They did not explore the consequences of these jumps for long-term dynamical evolution. Here we have demonstrated that jumps alter the maximum eccentricity $e_\mathrm{max}$ and secular period $P_\mathrm{K}$, sometimes by orders of magnitude (e.g., Figure \ref{fig:gr-scatter}) and can drastically change the time spent above a given eccentricity $T(> e)$ (e.g., Figure \ref{fig:gr-residence}).

\end{itemize}

The jumps in $j_z$ deserve further comment.  They are \textit{inevitable} because the allowed range of $\CK$ decreases with $\vert j_z \vert$ (\S\ref{sec:DA}); thus a jump that drives $\CK$ toward the edge of its range often cannot be accommodated at fixed $j_z$, so $\vert j_z \vert$ tends to \textit{decrease on average}. To our knowledge this is the first known \rev{test-particle} quadrupole-order, three-body mechanism that drives long-term evolution of $j_z$. {Previously, \rev{in the test-particle limit,} long-term evolution of $j_z$ was understood to require either octupole-order coupling (the EKL orbit-flip mechanism; e.g.,~\citealt{lithwick2011,katz2011,naoz2016,sidorenko2018,lei2022,klein2024a}), the addition of a fourth body in a quadruple configuration (e.g.,~\citealt{pejcha2013,hamers2017,fang2018,grishin2018quad,klein2023,klein2024d,klein2024c,liu2026}), or an aspherical cluster or Galactic tidal potential perturbing the triple \citep{petrovich2017,grishinperets2022}.}   Such evolution is important because $j_z$ is the main determinant of the eccentricity maximum $e_\mathrm{max}$. Thus, \textit{on average}
nonadiabatic jumps drive the binary towards ever more extreme orbital dynamics. 

In the past, SA dynamics was sometimes treated as a minor correction to DA dynamics, via approximations such as equations \eqref{eqn:bad_period_approx}--\eqref{eqn:bad_emax_approx} \citep{grishin2018,hamilton2019a}. Nonadiabatic jumps invalidate this approach: \rev{the resulting cycle-to-cycle variability is, to our knowledge, not captured (even statistically) by any current analytic or semi-analytic theory.}
Such variability means that \rev{a single, well-defined $e_\mathrm{max}$ and $P_\mathrm{K}$ per cycle --- the answers to questions (i) and (ii) posed in the Introduction ---} cease to be meaningful for systems residing in the `fourth quadrant' of Figure \ref{fig:quadrants}.  
We therefore conclude that, \rev{where both SA fluctuations and SRFs matter, the cycle-to-cycle $e_\mathrm{max}$, $e_\mathrm{min}$ and $P_\mathrm{K}$, and the secular evolution of $j_z$,} cannot be approximated by any currently-known DA or modified-DA theory: only brute-force integration of the SA (or $N$-body) equations of motion can be relied upon.\footnote{\rev{One might be able to use a \textit{fused} scheme, where DA is integrated ``away'' from high $e$ (for example as long as $e<0.99$), and at high-$e$ phases SA is used to account for the jump mechanism.}} \rev{(An upper bound on $e_\mathrm{max}$ over all initial conditions does admit an analytic estimate in the SA model \citep{huang2026}, as we discuss below.)}

This conclusion also sharpens how the DA-validity criterion \eqref{eq:da-criterion} should be applied in practice. In the compact-object--merger literature the DA criterion is conventionally written as $t_{\rm LK}(1-e_\mathrm{max}^2)^{1/2}\gtrsim \Pout$ \citep[e.g.,][]{liu2018,hamilton2021}, where $t_{\rm LK}$ stands in for the secular timescale; with our normalization $t_{\rm LK}=\Psec$, so this reads $\Pout/\Psec \lesssim j_{\rm min}^{\rm DA}$. Thus, DA is frequently assumed reliable whenever $\Pout/\Psec$ is merely \emph{somewhat} smaller than $j_{\rm min}^{\rm DA}$. A rigorous asymptotic statement of DA validity, however, calls for the strong inequality $\Pout/\Psec \ll j_{\rm min}^{\rm DA}$ (Eq. \eqref{eq:da-criterion}) rather than the marginal $\lesssim$. This might seem a pedantic distinction, but in fact it is crucially important. For instance, \citet{liu2018} partitioned their perturber mass--semimajor-axis plane of merging compact-object triples into adjacent `DA' and `SA' regions whose dividing curve is exactly $\Pout/\Psec = j_{\rm min}^{\rm DA}$ (their Figure~2). 
Our GR-only example (\S\ref{sec:gr-example}) would be deemed safe for DA according to this partitioning (see Eqs. \eqref{eq:gr_period_ratio} and \eqref{eq:gr-example-leaf}), but Figure \ref{fig:gr-evolution} makes clear that in fact, $\mathcal{O}(1)$ jumps in $\CK$, and secular evolution of $j_z$, both occur. \rev{While DA/SA discrepancies in $e_\mathrm{max}$ \textit{have} been reported in ZLK calculations including SRFs \citep{liu2018,huang2026}, they concern different comparisons from ours. \citet{ivanov2005,grishin2018} predict the peak eccentricity of a \emph{given} trajectory to be its double-averaged value plus a fluctuation $\delta e$ (equation \eqref{eqn:bad_emax_approx}), both set by that trajectory's initial conditions. \citet{huang2026} instead derive an \emph{upper bound} on $e_\mathrm{max}$, taken over all inner-orbit initial conditions, that is fixed by the SRF strengths and $e_2$ alone (independent of $j_z$ and $\CK$) and is higher than the DA maximum, obtained by balancing the maximal SA angular-momentum variation rate against the SRF precession rate. Even with \citet{huang2026}, then, no analytic prediction of $e_\mathrm{max}$ for a \emph{given} initial condition exists in the fourth quadrant, of the kind \citet{ivanov2005,grishin2018} provide in the adiabatic regime. Our result is distinct from both, and concerns the {sign} of the SRF's effect. At the DA level the precession always \emph{suppresses} $e_\mathrm{max}$ (or is so weak as to do nothing, as in the tidal example we show). By contrast, in every example we present, apsidal precession that suppresses (or does nothing to) the maximum eccentricity under double-averaging instead \emph{catalyzes} it under single-averaging, through discrete cycle-to-cycle jumps in $\CK$ that also reshape $e_\mathrm{min}$, $P_\mathrm{K}$, and the secular evolution of $j_z$. }

This may have significant astrophysical consequences, because  nonadiabatic  jumps lead to qualitatively different statistics for extreme-eccentricity passages than would have been predicted in their absence. 
In real systems, dissipative processes such as tides or GW emission can turn sufficiently high-eccentricity passages into migration or merger channels \citep{fabrycky2007,wen2003,miller2002,liu2017,liu2018,liu2019}. 
Thus, if we get those high-eccentricity statistics wrong, we might badly misestimate the observable rate of LIGO/Virgo sources, hot Jupiters, etc.
Population synthesis calculations of such processes that relied on formulae like \eqref{eqn:bad_period_approx}-\eqref{eqn:bad_emax_approx} may therefore need to be revisited.

Our mechanism should not be confused with the relativistic `precession resonance' of \citet{kuntz2022}, in which the GR apsidal precession of the inner orbit becomes commensurate with the outer orbital frequency and the eccentricity is resonantly excited to a bounded ceiling. The two effects differ fundamentally. In \citet{kuntz2022}'s resonance, the rapid GR precession occurs at low eccentricity, because the binary radiates gravitational waves and thus shrinks its semimajor axis. In this regime SRFs are strong enough to  dominate the secular dynamics ($\epsGR \sim \pi \Psec/\Pout \gg 1$, so ZLK cycles are fully suppressed). By contrast, our jumps are purely conservative --- the inner semimajor axis is fixed and no dissipation enters --- and our SRFs have $\varepsilon_i \ll 1$, so ZLK cycles operate. While our systems \textit{do} reach very fast precession rates that become commensurate with the outer orbital period, they do so only transiently during a \rev{high-$e$ phase}, and for the rest of their evolution the apsidal precession is much slower than the outer orbital frequency. The `resonance' condition in our case is satisfied far too briefly for \citet{kuntz2022}'s adiabatic capture mechanism to operate.





We studied only the test-particle quadrupole version of ZLK dynamics in this paper, purely for simplicity, in order to exhibit nonadiabatic jumps in the cleanest possible setting.  Relaxing any of these assumptions will only make the dynamics more extreme. 
At octupole order, for instance, a
nonzero outer eccentricity combined with unequal inner masses can drive secular evolution of $j_z$ even at the DA level
\citep{katz2011,lithwick2011,naoz2016,klein2024a}.  This evolution is strongly controlled by the 
sign and value
of $\CK$: librating trajectories ($\CK < 0$) undergo little or no secular $j_z$ evolution, whereas for rotating ones ($\CK > 0$) it can be highly efficient \revcite{katz2011,klein2024a,klein2026}.
Since nonadiabatic jumps can flip the sign of $\CK$, they can naturally trigger or quench this octupolar transport. The interplay between nonadiabatic jumps and octupole-order ZLK dynamics is therefore a natural
extension of this study, which we pursue in a companion \rev{work} \citep{klein_hamilton_in_prep}.
Similarly, there might be some non-trivial interplay between nonadiabatic jumps discussed here and the very long-term modulation of ZLK cycles due to an accumulation of SA terms that is encapsulated in the Brown Hamiltonian \citep{tremaine2023}.

Finally, we showed some of the \textit{consequences} of nonadiabatic jumps in this paper but we did not study in detail their \textit{causes}.  
Some progress was made along these lines by \citetalias{hamilton2024} (with GR as the only SRF), but there is much more to do. In particular it appears from numerical experiments that the microscopic details of the jump behavior and conditions do depend on the SRF in question. 
\rev{We emphasize that individual jumps are \textit{not} random: for a given parent cycle the jump in $\CK$ is an ordered, reproducible function of the initial outer phase (mean anomaly $M_2$), and the long-term evolution only \emph{looks} irregular because successive cycles sample almost the entire range of $M_2$ (Appendix~\ref{sec:SA_Validity}). We do not, however, have a closed-form expression for this single-jump dependence, which we leave to future work.} \rev{A related analytic programme has been carried out for the cluster-driven analogue of this problem by \citet{ivanovpolnarev2025}, who matched the near-adiabatic evolution across each high-$e$ episode to construct a cycle-to-cycle map.}

Finally, we did not develop any theory capable of predicting the jumps' statistics. Since very long-term behavior is sensitive to changes in choices of initial phases, the exact approximations used, etc. (see Appendix \ref{sec:SA_Validity}), in the end it may only be a statistical description that
is physically meaningful. 
We defer a careful study to future work.

\section{Summary}
\label{sec:summary}

In this paper we have studied the classic problem of hierarchical triple dynamics in the test-particle quadrupole limit. In particular, we studied the resulting von Zeipel-Lidov-Kozai (ZLK) oscillations in the single-averaged (SA) approximation, \textit{including} precession due to short-range forces (SRFs). 
Our  key  findings can be summarized as follows.
\begin{itemize}

\item Systems that break the double averaging (DA) approximation (so the validity criterion \eqref{eq:da-criterion} is violated) \textit{and} are subject to apsidal precession from SRFs (i.e., those sitting in the fourth quadrant of Figure \ref{fig:quadrants}) can undergo nonadiabatic jumps in their approximate `invariant' labels $j_z$ and $C_\mathrm{K}$.

\item These jumps can lead to much more rapid exploration of phase space, including more extreme eccentricity behavior, than would otherwise be expected. They also provide, to our knowledge, the only known \rev{test-particle} quadrupole-order, three-body mechanism that drives a net, long-term evolution of the angular-momentum component $j_z$.

\item No known semi-analytic/modified-DA theory is capable of usefully predicting \rev{the cycle-to-cycle distribution of $e_\mathrm{max}$ and $P_\mathrm{K}$, or the secular evolution of $j_z$, that these jumps produce.} \rev{For these cycle-to-cycle quantities,} only numerical integration of the SA or $N$-body equations of motion can be trusted.

\item {These effects} may have significant astrophysical consequences. For instance, time spent above a particular eccentricity $T(>e)$ can be orders of magnitude different in the presence of nonadiabatic jumps. Various population synthesis calculations may need to be revisited in light of this fact. 

\end{itemize}
In a companion \rev{work} we show how ZLK cycles subject to nonadiabatic jumps become even more extreme when octupolar effects are important.

\begin{acknowledgments}
We thank the anonymous referee for constructive comments.
We thank Scott Tremaine for useful discussions.
We thank Xiumin Huang for fruitful comments.
We also thank Evgeni Grishin, Pavel Ivanov, Boaz Katz, \rev{Hanlun Lei, Crist\'obal Petrovich,} Yubo Su and Roman Rafikov for valuable feedback.
C.H. was partly supported by the John N. Bahcall Fellowship Fund at the Institute for Advanced Study.
\end{acknowledgments}


\bibliographystyle{aasjournal}
\bibliography{papers}{}

\appendix 

\section{Initial conditions and numerical integration}
\label{sec:specifying_ICs}

In \S\ref{sec:numerical} we quoted the initial conditions of each example as standard orbital elements ($e$, $\omega$, $i$, $\Omega$, and so on). That is not how we actually set up or run the integrations. In practice we work entirely in dimensionless variables: we fix each system by a small set of dimensionless numbers, build the initial $\mathbf{j}$ and $\mathbf{e}$ vectors from that set, integrate the dimensionless equations of motion, and only afterwards, when we want a concrete astrophysical system, do we convert to physical units.

Up to an overall timescale, the dynamics of any system we study is fixed by the nine dimensionless numbers
\begin{equation}
j_z,\ \CK,\ \Omega,\ e_2,\ \epsGR,\ \epsRot,\ \epsTide,\ \Pout/\Psec,\ M_2.
\label{eq:dimensionless-set}
\end{equation}
All systems that share this set evolve identically as functions of $\tau=t/\Psec$. We specify these numbers first, and everything else follows from them: the pair $(j_z,\CK)$ labels the underlying no-SRF DA quadrupole cycle; $e_2$ and $\Omega$ orient the inner orbit relative to the outer apse; $(\epsGR,\epsRot,\epsTide)$ set the SRF strengths through Eqs.~\eqref{eq:epsGR}--\eqref{eq:epsTide}; and $\Pout/\Psec$ and the initial outer mean anomaly $M_2$ fix the phase of the outer orbit, which single-averaging retains.\footnote{In DA, $\Omega$ is a gauge angle, since the potential is axisymmetric about the outer angular-momentum axis; the same holds in SA when $e_2=0$. Only for SA with $e_2>0$ does the outer eccentricity vector pick out a preferred apsidal direction, which makes $\Omega$ a genuine initial condition.}

From this set we build the initial state in three steps. First, from $(j_z,\CK)$ we compute the minimum eccentricity $\emin$ of the no-SRF DA quadrupole cycle \citep{antognini2015,basha2025,kinoshita2007,vashkovyak1999} and start every integration there, $e=\emin$, so that each run begins at a well-defined phase of its cycle. Second, we fix the apsidal branch from the sign of $\CK$: we take $\omega=\pi/2$ for librating cycles ($\CK<0$) and $\omega=0$ for rotating cycles ($\CK>0$). Third, we orient the orbit in the outer frame with $\Omega$. These three choices determine the initial $\mathbf{j}$ and $\mathbf{e}$.
We then integrate the Milankovitch equations \eqref{eq:milankovitch-j}--\eqref{eq:milankovitch-e} with the dimensionless potential $\phi=\phi_{\rm quad}+\phi_{\rm SRF}$, taking the quadrupole part to be either the single-averaged potential $\phi_{\rm quad}^{\rm SA}$ \eqref{eq:phi-SA-quad} or the double-averaged potential $\phi_{\rm quad}^{\rm DA}$ \eqref{eq:phi-quad} as appropriate, forward in dimensionless time $\tau=t/\Psec$.

Finally, to display a system in physical units we attach a single dimensional scale (equivalently, the component masses and the two semimajor axes), whereupon the dimensionless set \eqref{eq:dimensionless-set} maps onto dimensional initial conditions. In the next subsection we list such dimensional realizations for each example.

\subsection{Dimensional realizations of the numerical examples}
\label{sec:dimensional-realizations}

The three numerical examples of \S\ref{sec:numerical} are specified by their
dimensionless parameter sets (Appendix~\ref{sec:specifying_ICs}); the dynamics depend
only on these dimensionless numbers, so each parameter set corresponds to a whole family
of dimensional systems related by scaling, all evolving identically as functions
of $\tau=t/\Psec$. For reproducibility, in Table~\ref{tab:dimensional} we give one
explicit, high-precision dimensional realization of each example, i.e., we give masses,
semimajor axes, eccentricities, body parameters and orientation angles one
would supply to a secular or $N$-body integrator. The inner orbit starts at its
DA eccentricity minimum, $e=\emin(j_z,\CK)$, with $\omega=0$ (rotating
branch, $\CK>0$; Appendix~\ref{sec:specifying_ICs}); for these parameter sets
($\CK=10^{-6}$) this is $e=\sqrt{\CK}=1.0\times10^{-3}$, and the round value
$e=0$ quoted in the main text is shorthand for this small but nonzero minimum.
{For the GR example (\S\ref{sec:gr-example}) we list three realizations of the
\emph{same} dimensionless parameter set, all orbiting an $m_2=10^{7}\,M_\odot$ SMBH on an
$e_2=0.8$ orbit: a binary neutron star (NS+NS), a $10+10\,M_\odot$ BBH, and the
$30+30\,M_\odot$ BBH shown in the main text.}  For the remaining two examples (tides and rotation, \S\S\ref{sec:tides-example}-\ref{sec:rotation-example}) we show only one realization.

\begin{deluxetable*}{l ccc cc}
\tablecaption{Dimensional realizations of the three numerical examples of
\S\ref{sec:numerical}. The three GR columns share one dimensionless parameter set and
differ only by an overall scaling. Inner-orbit angles follow from
$(j_z,\CK)$ via Appendix~\ref{sec:specifying_ICs}.\label{tab:dimensional}}
\tablehead{
\colhead{} & \colhead{GR: NS+NS} & \colhead{GR: $10{+}10\,M_\odot$} &
\colhead{GR: $30{+}30\,M_\odot$} & \colhead{Tides:} & \colhead{Rotation:}\\
\colhead{} & \colhead{} & \colhead{} & \colhead{(shown)} &
\colhead{Sun--Neptune} & \colhead{Sun--Jupiter}
}
\startdata
\cutinhead{Bodies}
$m_0\;[M_\odot]$        & $1.4$   & $10$    & $30$    & $1$                  & $1$ \\
$m_1\;[M_\odot]$        & $1.4$   & $10$    & $30$    & $5.150\times10^{-5}$ ($M_{\rm Nep}$) & $9.546\times10^{-4}$ ($M_{\rm Jup}$) \\
$m_2\;[M_\odot]$        & $10^{7}$ & $10^{7}$ & $10^{7}$ & $0.642$            & $0.20$ \\
$R_1\;[\mathrm{AU}]$    & \nodata & \nodata & \nodata & $1.646\times10^{-4}$ ($R_{\rm Nep}$) & $4.778\times10^{-4}$ ($R_{\rm Jup}$) \\
\rev{$k_{2,1}$}         & \nodata & \nodata & \nodata & $0.127$              & \nodata \\
\rev{$k_{q,1}$}         & \nodata & \nodata & \nodata & \nodata              & \rev{$0.17$} \\
$P_{\rm spin}$          & \nodata & \nodata & \nodata & \nodata              & $9.93\,$hr \\
\cutinhead{Orbit (inner / outer)}
$a\;[\mathrm{AU}]$    & $0.763$ & $5.45$  & $16.34$ & $11.94$              & $7$ \\
$a_2\;[\mathrm{AU}]$    & $1.449\times10^{4}$ & $5.373\times10^{4}$ & $1.118\times10^{5}$ & $238.9$ & $100$ \\
$a_2\;[\mathrm{pc}]$    & $0.0703$ & $0.260$ & $0.542$ & \nodata             & \nodata \\
$e\;(=\emin)$         & $1.0\times10^{-3}$ & $1.0\times10^{-3}$ & $1.0\times10^{-3}$ & $1.0\times10^{-3}$ & $1.0\times10^{-3}$ \\
$e_2$                   & $0.8$   & $0.8$   & $0.8$   & $0$                  & $0$ \\
$i\;[\deg]$           & $89.4$  & $89.4$  & $89.4$  & $88.6$               & $89.4$ \\
$\omega\;[\deg]$      & $0$     & $0$     & $0$     & $0$                  & $0$ \\
$\Omega\;[\deg]$      & $180$   & $180$   & $180$   & $180$                & $180$ \\
$M_2\;[\deg]$           & $278.6$ & $278.6$ & $278.6$ & $117.3$              & $349.5$ \\
\cutinhead{Periods}
$P_{\rm in}\;[\mathrm{yr}]$ & $0.398$ & $2.845$ & $8.527$ & $41.3$           & $18.51$ \\
$\Pout\;[\mathrm{yr}]$  & $551$   & $3939$  & $1.181\times10^{4}$ & $2881$   & $912$ \\
$\Psec\;[\mathrm{Myr}]$ & $0.0262$ & $0.187$ & $0.562$ & $0.082$            & $0.0430$ \\
\cutinhead{Dimensionless parameter set}
$j_z$                   & $0.010$ & $0.010$ & $0.010$ & $0.025$             & $0.010$ \\
$\CK$                   & $10^{-6}$ & $10^{-6}$ & $10^{-6}$ & $10^{-6}$     & $10^{-6}$ \\
$\Pout/\Psec$           & $0.021$ & $0.021$ & $0.021$ & $0.035$             & $0.021$ \\
$\epsGR$                & $0.0451$ & $0.0451$ & $0.0451$ & $0$              & $0$ \\
$\epsTide$              & $0$     & $0$     & $0$     & $3.08\times10^{-16}$ & $0$ \\
$\epsRot$               & $0$     & $0$     & $0$     & $0$                  & $1.0\times10^{-6}$ \\
\cutinhead{\rev{Hierarchical stability}}
\rev{$a_2(1-e_2)/a$}                     & \rev{$3.80\times10^{3}$} & \rev{$1.97\times10^{3}$} & \rev{$1.37\times10^{3}$} & \rev{$20.0$} & \rev{$14.3$} \\
\enddata
\end{deluxetable*}

\rev{The configurations in Table~\ref{tab:dimensional} are all dynamically stable hierarchical triples, and they remain so throughout their high-eccentricity evolution. Hierarchical stability is set by the ratio of the outer pericenter to the inner semimajor axis, $a_2(1-e_2)/a$, together with the masses, $e_2$, and the mutual inclination; it does \emph{not} depend on the inner eccentricity, because in the conservative quadrupole test-particle limit $a$ is conserved while only $e$ evolves. For the two examples with a stellar-mass tertiary (tides and rotation), i.e.\ an order-unity outer mass ratio $m_2/(m_0+m_1)$, the \citet{mardling2001} criterion requires $a_2(1-e_2)/a \gtrsim 2.8\,[(1+m_2/(m_0+m_1))(1+e_2)(1-e_2)^{-1/2}]^{2/5}(1-0.3\,i/180^\circ)$, which our examples satisfy by factors of \rev{$6.9$ (tides) and $5.6$ (rotation)}. For the three SMBH realizations, the relevant condition is that the inner binary not be tidally disrupted by the tertiary, i.e.,\ that its apocenter stay within the Hill radius of the inner binary about the SMBH, $a(1+e)<r_{\rm H}=a_2(1-e_2)[(m_0+m_1)/3m_2]^{1/3}$ (evaluated at outer pericenter). This is satisfied by a factor ${\sim}9$ even as $e\to1$. Crucially, the high-eccentricity excursions we study do not endanger the hierarchy: since $a$ is fixed, the inner apocenter never exceeds $2a$, far interior to the outer pericenter, so the orbits never cross and the secular treatment remains self-consistent. What ultimately limits the eccentricity growth is instead the inner binary's own short-range physics (gravitational-wave or tidal dissipation), which converts sufficiently deep pericenter passages into mergers or tidal disruptions rather than three-body instabilities (\S\ref{sec:discussion}).}

\section{Validity of the single-averaging and test-particle approximations}
\label{sec:SA_Validity}

\begin{figure}
    \centering
    \includegraphics[width=0.8\columnwidth]{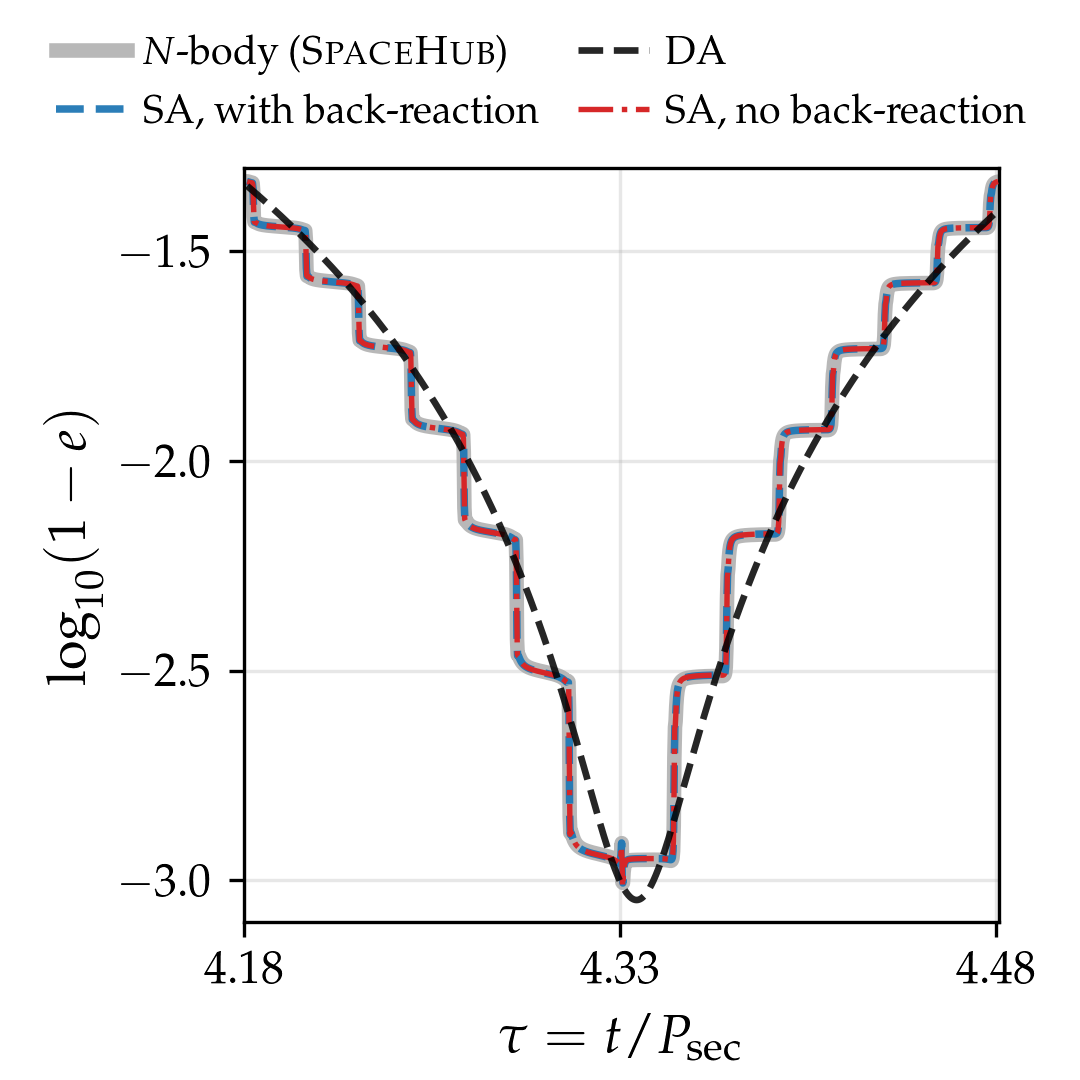}
    \caption{The first eccentricity peak from Figure \ref{fig:gr-evolution}, integrated with multiple methods. The \rev{red dot-dashed} and black dashed lines reproduce the SA and DA results from Figure \ref{fig:gr-evolution} respectively. The blue dashed line shows the SA integration without the test-particle approximation (i.e., including the back-reaction onto the outer orbit), and the thick gray line shows the results of a direct $N$-body integration  (\textsc{SpaceHub}).}
    \label{fig:nbody-zoom}
\end{figure}
\begin{figure}
    \centering
    \includegraphics[width=0.49\textwidth]{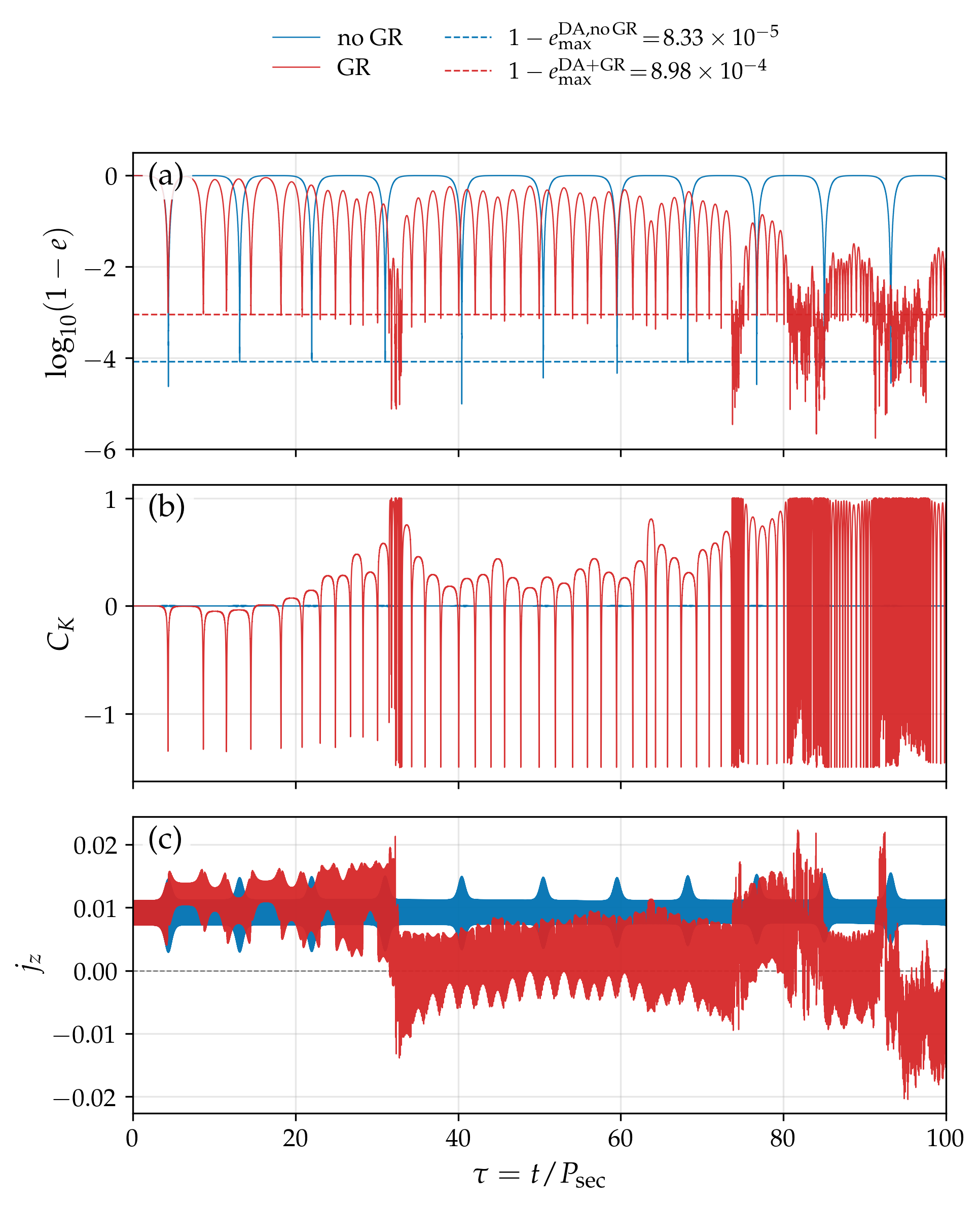}
    \caption{\rev{Direct $N$-body (\textsc{SpaceHub}) evolution of the same BBH--SMBH triple as in Figure~\ref{fig:gr-evolution} (\S\ref{sec:gr-example}), integrated with 1pN GR precession switched on (red; the only SRF in this system) and off (blue), with all other parameters held fixed. Panels show (a)~$\log_{10}(1-e)$, (b)~$\CK$ and (c)~$j_z$ versus $\tau = t/\Psec$; the horizontal dashed lines in panel~(a) mark the double-averaged $\emax$ ceilings (no-GR, blue; $+$GR, red), and the horizontal dashed line in panel~(c) marks $j_z=0$.}}
    \label{fig:nbody-evolution}
\end{figure}
\begin{figure}
    \centering
    \includegraphics[width=0.95\columnwidth]{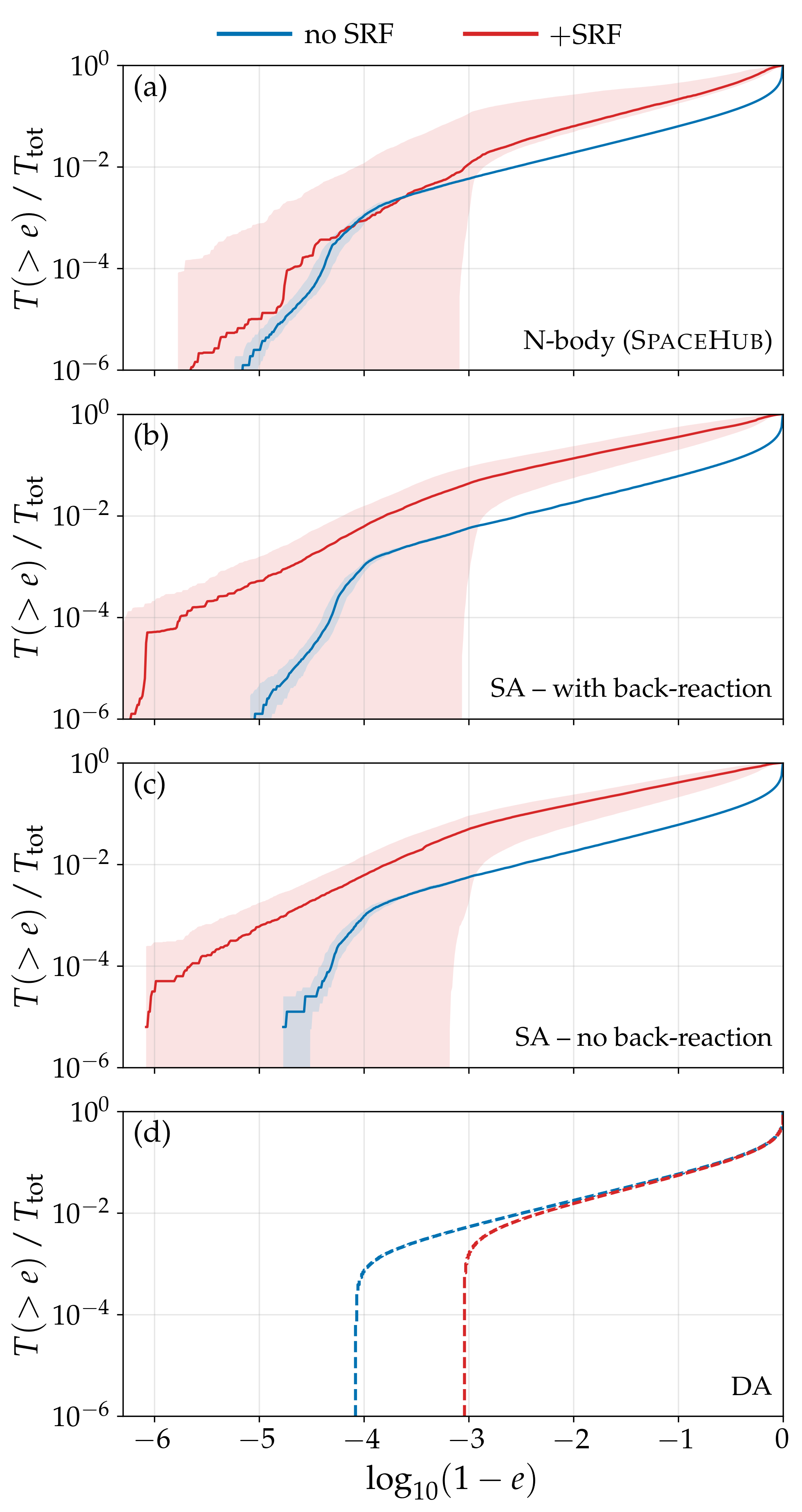}
    \caption{Cumulative residence time $T(>e)/T_\mathrm{tot}$ versus $\log_{10}(1-e)$ for the BBH--SMBH triple of \S\ref{sec:gr-example} (Figures \ref{fig:gr-evolution}--\ref{fig:gr-residence}), computed using the same four different methods used for Figure~\ref{fig:nbody-zoom}. Panels show (a)~direct $N$-body (\textsc{SpaceHub}), (b)~SA with back-reaction on the outer orbit, (c)~SA in the test-particle limit (no back-reaction), and (d)~DA. Colors are the same as in Figure \ref{fig:gr-residence}. The initial conditions are those of the $30+30\,M_\odot$ BBH listed in Table~\ref{tab:dimensional}.}
    \label{fig:nbody}
\end{figure}

Throughout the main text we relied on two simplifications of the underlying few-body dynamics: (i) single-averaging (SA) of the inner orbit (ignoring all fluctuations on the timescale of the inner orbital period $P_\mathrm{in}$), and (ii) the test-particle approximation (so the outer orbit is perfectly Keplerian and does not feel any back-reaction from the inner binary dynamics). The latter approximation is what allowed the problem to be cast in fully dimensionless form (Appendix \ref{sec:specifying_ICs}). 
Both of these approximations can break down in certain circumstances \citep{antonini2014,anderson2016}. However, we checked that the qualitative phenomenology established in \S\ref{sec:numerical} (jumps in $\CK$, the SRF-induced secular evolution of the adiabatic invariants $\CK$ and $j_z$, and the resulting SRF-catalyzed excursions to extreme eccentricity, all captured by the cumulative residence-time diagnostic) is robust to these approximations. Here we give some concrete examples.

In Figure \ref{fig:nbody-zoom} we zoom in on the first eccentricity maximum ($\tau\approx 4.33$) of the trajectory shown in Figure \ref{fig:gr-evolution}, which we recall is for a BBH--SMBH triple (see \S\ref{sec:gr-example}).  We show the results of integrating this system using multiple different methods. The \rev{red dot-dashed} and black dashed lines reproduce the SA and DA results from Figure \ref{fig:gr-evolution} respectively. The blue dashed line shows the SA integration without the test-particle approximation (i.e., including the back-reaction onto the outer orbit) \citep{mangipudi2022}, and the thick gray line shows the results of a direct $N$-body integration (no averaging at all) using \textsc{SpaceHub} \citep{wang2021_spacehub}. We see that the two SA integrations and the $N$-body integration produce results that are indistinguishable, giving us confidence in our approach.

\rev{Figure~\ref{fig:nbody-zoom} establishes this agreement for the first eccentricity peak. To test whether the qualitative mechanism itself --- the jumps in $\CK$ and the secular drift of $j_z$ --- survives a direct integration over many secular cycles, in Figure~\ref{fig:nbody-evolution} we show the full $N$-body evolution of the same BBH--SMBH triple, integrated with \textsc{SpaceHub}, with GR (the only SRF present in this system) switched on and off. We deliberately do not overlay the DA and SA trajectories here: as discussed in the next paragraph, individual trajectories decohere in phase after the first few cycles, so a cycle-by-cycle overlay of different methods is uninformative beyond the first peak. What \emph{is} robust is the mechanism itself. With GR off, $\CK$ and $j_z$ stay regular; with GR on, the $N$-body integration develops the same $\mathcal{O}(1)$ jumps in $\CK$ and secular drift of $j_z$ that we found in the single-averaged treatment (Figure~\ref{fig:gr-evolution}). The SRF-catalyzed route to extreme eccentricity is therefore present in the full $N$-body problem, and is not an artifact of single-averaging.}

However, we are concerned not only with the dynamics of a single eccentricity peak, but also with the statistics of long-term evolution over many secular cycles. As is well-known in planetary dynamics \citep{laskar1989,sussman1992}, tiny changes in the approximations used (and even in numerical choices like the exact timestep) can, over long enough timescales, creep in to the dynamics and produce, e.g., phase offsets between the results of different integrations. Since nonadiabatic jumps are sensitive to orbital phases, these offsets can drive altogether different orbital behavior. \rev{We stress that this sensitivity is a feature of the secular dynamics, not a symptom of incipient dynamical instability: by the Mardling--Aarseth \citep{mardling2001} and Hill stability criteria the configurations are hierarchically stable, independently of how high the inner eccentricity is driven (see the hierarchical-stability discussion in Appendix~\ref{sec:dimensional-realizations}), and the same jumps and high-$e$ excursions appear in direct $N$-body integration (Figure~\ref{fig:nbody-evolution}).} Thus we should not rely upon \textit{any} integration method to be exact: all one can ultimately rely upon is the \textit{statistics} that these methods produce.

To study this, in Figure \ref{fig:nbody} we recreate the residence time plot for the BBH--SMBH triple of \S\ref{sec:gr-example} (cf. Figure \ref{fig:gr-residence}) using four different dynamical treatments. The curves in panel (c) are identical to those in Figure \ref{fig:gr-residence}, showing the result for SA dynamics without any back-reaction onto the outer orbit. 
Panel (b) re-integrates the same $50$ realizations with the  back-reaction restored (i.e., without the test-particle approximation).
Panel (d) shows the realization-independent DA reference curves from Figure \ref{fig:gr-residence}. Panel (a) re-integrates them with the $N$-body code.
The residence time diagnostic is useful because it combines two kinds of averaging: it integrates over the full time evolution of each trajectory \textit{and} aggregates over the ensemble of outer mean-anomaly realizations $M_2$.

Three points are immediate. First, panels (a), (b) and (c) agree at the qualitative level across the plotted eccentricity range: the SRFs produce the same characteristic enhancement of the high-$e$ residence tail, the $+$SRF (red) median crosses to the left of the no-SRF (blue) median at the same $\log_{10}(1-e)\simeq -3$, and the realization-to-realization spread covers a comparable few-decade band. Secondly, the close agreement between panels (b) and (c) shows that the inner-on-outer back-reaction has negligible effect on the residence statistics for this system, validating the test-particle approximation. Thirdly, the DA panel (d) is qualitatively distinct from the other three: the residence curves terminate abruptly at the analytic DA $e_\mathrm{max}$ values and have the $+$SRF branch to the \emph{right} of the no-SRF branch (the opposite of the SA/$N$-body ordering).  The main-text story that SRFs catalyze, rather than suppress, excursions to extreme eccentricity is therefore an SA feature that survives both the back-reaction and the full $N$-body integration, but is absent in the DA limit.

\section{Onset of jumps below the naive threshold}
\label{sec:onset-belowthreshold}

\begin{figure}
\centering
\includegraphics[width=\columnwidth]{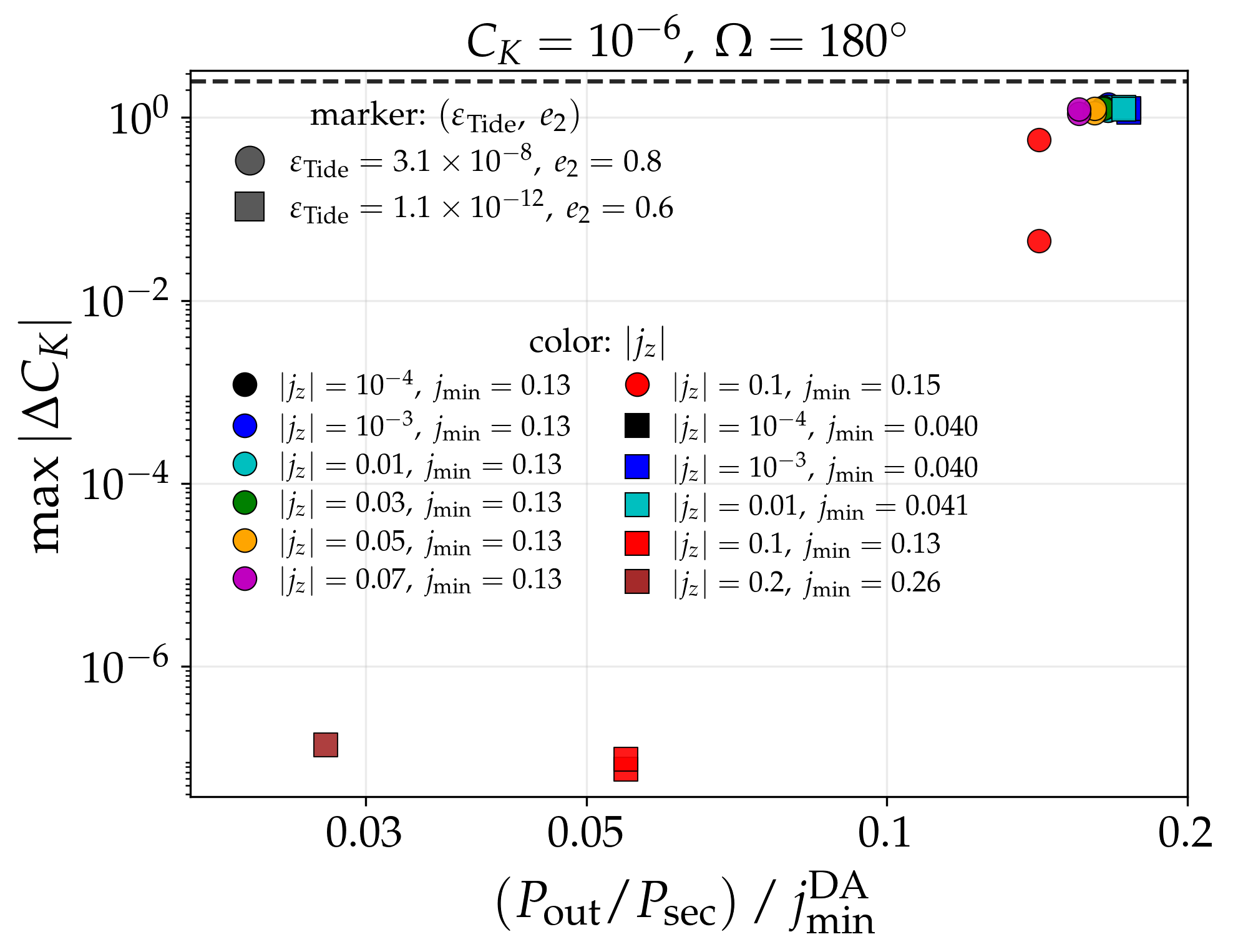}
\caption{\rev{Maximum single-cycle jump in the Kozai constant, $\max|\Delta\CK|$ \eqref{eq:maxdck}, versus $(\Pout/\Psec)/j_\mathrm{min}^\mathrm{DA}$ for two strong-tide families (circles and squares; tidal strength $\epsTide$ and outer eccentricity $e_2$ as labelled), each scanned over the inner-orbit inclination through $j_z$ (both signs). Color encodes $|j_z|$ and the legend lists the corresponding DA minimum angular momentum $j_\mathrm{min}^\mathrm{DA}$; all systems have $\CK=10^{-6}$, $\Omega=180^\circ$, and pure tidal short-range forcing ($\epsGR=\epsRot=0$), isolating the tidal-bulge effect. The dashed line is $|\Delta\CK|=5/2$.}}
\label{fig:onset-strongtide}
\end{figure}

\rev{The arguments made in the main text, and the results of Figure~\ref{fig:onset}, suggest the onset of nonadiabatic $\CK$ jumps will occur at $\Pout/\Psec\simeq j_\mathrm{min}^\mathrm{DA}$.
However, as emphasized several times above, this threshold is only a heuristic: the true onset depends on the full location in the dimensionless phase space \eqref{eq:dimensionless-set}. Figure~\ref{fig:onset-strongtide} shows two families of systems for which the jumps set in well before this threshold is reached. From the tidally dominated Cases~6 and~9 of \citet[][Table~1]{liu2015} we adopt the two realistic ingredients --- the outer eccentricity $e_2$ and the tidal strength $\epsTide$ --- and sweep $\Pout/\Psec$ as in Figure~\ref{fig:onset}, so the nominal $\Pout/\Psec$ of each Table~1 system is only one point on its sweep. In both cases a Jupiter ($m_1=1\,M_{\rm Jup}$, $R_1=R_{\rm Jup}$, Love number $k_2=0.37$) orbits a solar-mass primary ($m_0=1\,M_\odot$) and is perturbed by a tertiary at $a_2=100$~AU: Case~6 has $m_2=1\,M_\odot$, $a=1$~AU and $e_2=0.8$, giving $\epsTide\approx3\times10^{-8}$; Case~9 has $m_2=0.04\,M_\odot$, $a=6$~AU and $e_2=0.6$, giving $\epsTide\approx1\times10^{-12}$. Realistically, planetary systems with such eccentric outer orbits would warrant an octupole-level treatment, but here we switch the octupole off by hand precisely to show that the nonadiabatic jumps already arise at quadrupole order. Similarly, the only SRF retained here is from the tidal bulge ($\epsGR=\epsRot=0$), though in reality other SRFs would be present too. We see from Figure \ref{fig:onset-strongtide} that large nonadiabatic jumps occur even at $(\Pout/\Psec)/j_\mathrm{min}^\mathrm{DA}\simeq 0.1$. This reinforces our warning that the criterion \eqref{eq:da-criterion} must not be applied too loosely: a system with $\Pout/\Psec$ only modestly below $j_\mathrm{min}^\mathrm{DA}$ can already be strongly nonadiabatic.}

\end{document}